 \documentclass[preprint,prd,tightenlines, superscriptaddress]{revtex4-1}

\usepackage{graphicx} 
\usepackage{dcolumn}  
\usepackage{colordvi}
\usepackage{color}
\usepackage{epstopdf}
\usepackage{pstricks}
\usepackage{amssymb}
\usepackage{url}
\graphicspath{{ps}}
\usepackage{hyperref}
\usepackage{tabularx}
\usepackage{multirow}
\usepackage{units}
\usepackage{upgreek}
\usepackage{siunitx}
\usepackage{hyphenat}
\usepackage{subfigure}
\usepackage[italic]{hepnames}

\newcommand{\CP}{\ensuremath{C\hspace{-0.13em}P}\xspace}

\renewcommand{\PBzero}{\ensuremath{\HepParticle{\PB}{}{}^0}\xspace}
\renewcommand{\APBzero}{\ensuremath{\HepParticle{\APB}{}{}^0}\xspace}

\renewcommand{\APDzero}{\ensuremath{\HepParticle{\APD}{}{}^0}\xspace}
\renewcommand{\Pgpz}{\ensuremath{\HepParticle{\Pgp}{}{}^0}\xspace}

\renewcommand{\PKzS}{\ensuremath{\HepParticle{\PK}{}{}^0_{\rm S}}\xspace}

\newcolumntype{Y}{>{\centering\arraybackslash}X}

\def\mbc{\ensuremath{M_{{\rm bc}}}\xspace}

\def\de{\ensuremath{\Delta E}\xspace}

\def\q{\ensuremath{q}\xspace}

\def\r{\ensuremath{r}\xspace}
\def\w{\ensuremath{w}\xspace}

\def\wr{\ensuremath{w_{{r}}}\xspace}

\def\er{\ensuremath{\epsilon_{{r}}}\xspace}

\def\akzpiz{\ensuremath{\mathcal{A}_{{K^0\pi^0}}}\xspace}

\def\ikpi{\ensuremath{I_{{K\pi}}}\xspace}

\RequirePackage{xspace}

\def\belletwo {Belle~II}

\def\q     {\ensuremath{q}\xspace}

\def\piz   {\ensuremath{\pi^0}\xspace}

\def\pip   {\ensuremath{\pi^+}\xspace}
\def\pim   {\ensuremath{\pi^-}\xspace}

\def\Kbar  {\kern 0.2em\overline{\kern -0.2em K}{}\xspace}

\def\Kz    {\ensuremath{K^0}\xspace}
\def\Kzb   {\ensuremath{\Kbar^0}\xspace}
\def\KzKzb {\ensuremath{\Kz \kern -0.16em \Kzb}\xspace}
\def\Kp    {\ensuremath{K^+}\xspace}
\def\Km    {\ensuremath{K^-}\xspace}

\def\KpKm  {\ensuremath{\Kp \kern -0.16em \Km}\xspace}
\def\KS    {\ensuremath{K^0_{\scriptscriptstyle S}}\xspace}

\def\Dbar    {\kern 0.2em\overline{\kern -0.2em D}{}\xspace}

\def\Dz      {\ensuremath{D^0}\xspace}
\def\Dzb     {\ensuremath{\Dbar^0}\xspace}
\def\DzDzb   {\ensuremath{\Dz {\kern -0.16em \Dzb}}\xspace}
\def\Dp      {\ensuremath{D^+}\xspace}
\def\Dm      {\ensuremath{D^-}\xspace}

\def\DpDm    {\ensuremath{\Dp {\kern -0.16em \Dm}}\xspace}

\def\Bbar    {\kern 0.18em\overline{\kern -0.18em B}{}\xspace}

\def\Bz      {\ensuremath{B^0}\xspace}
\def\Bzb     {\ensuremath{\Bbar^0}\xspace}
\def\BzBzb   {\ensuremath{\Bz {\kern -0.16em \Bzb}}\xspace}
\def\Bu      {\ensuremath{B^+}\xspace}
\def\Bub     {\ensuremath{B^-}\xspace}
\def\Bp      {\ensuremath{\Bu}\xspace}

\def\BpBm    {\ensuremath{\Bu {\kern -0.16em \Bub}}\xspace}

\mathchardef\Upsilon="7107
\def\Y#1S{\ensuremath{\Upsilon{(#1S)}}\xspace}

\mathchardef\Deltares="7101
\mathchardef\Xi="7104
\mathchardef\Lambda="7103
\mathchardef\Sigma="7106
\mathchardef\Omega="710A

\def\Deltabar{\kern 0.25em\overline{\kern -0.25em \Deltares}{}\xspace}
\def\Lbar{\kern 0.2em\overline{\kern -0.2em\Lambda\kern 0.05em}\kern-0.05em{}\xspace}
\def\Sigbar{\kern 0.2em\overline{\kern -0.2em \Sigma}{}\xspace}
\def\Xibar{\kern 0.2em\overline{\kern -0.2em \Xi}{}\xspace}
\def\Obar{\kern 0.2em\overline{\kern -0.2em \Omega}{}\xspace}
\def\Nbar{\kern 0.2em\overline{\kern -0.2em N}{}\xspace}
\def\Xb{\kern 0.2em\overline{\kern -0.2em X}{}\xspace}

\newcommand{\tev}{\ensuremath{\mathrm{\,Te\kern -0.1em V}}\xspace}
\newcommand{\gev}{\ensuremath{\mathrm{\,Ge\kern -0.1em V}}\xspace}
\newcommand{\mev}{\ensuremath{\mathrm{\,Me\kern -0.1em V}}\xspace}
\newcommand{\kev}{\ensuremath{\mathrm{\,ke\kern -0.1em V}}\xspace}
\newcommand{\ev}{\ensuremath{\mathrm{\,e\kern -0.1em V}}\xspace}
\newcommand{\gevc}{\ensuremath{{\mathrm{\,Ge\kern -0.1em V\!/}c}}\xspace}
\newcommand{\mevc}{\ensuremath{{\mathrm{\,Me\kern -0.1em V\!/}c}}\xspace}
\newcommand{\gevcc}{\ensuremath{{\mathrm{\,Ge\kern -0.1em V\!/}c^2}}\xspace}
\newcommand{\mevcc}{\ensuremath{{\mathrm{\,Me\kern -0.1em V\!/}c^2}}\xspace}

\def\mus  {\ensuremath{\rm \,\mus}\xspace}

\def\mus        {\ensuremath{\,\mu{\rm s}}\xspace}    

\def\to                 {\ensuremath{\rightarrow}\xspace}

\def\gsim{{~\raise.15em\hbox{$>$}\kern-.85em
          \lower.35em\hbox{$\sim$}~}\xspace}
\def\lsim{{~\raise.15em\hbox{$<$}\kern-.85em
          \lower.35em\hbox{$\sim$}~}\xspace}

\begin{document}

\def\belletwo {\it {Belle~II}}

\clubpenalty = 10000  
\widowpenalty = 10000 

\vspace*{-3\baselineskip}
\resizebox{!}{3cm}{\includegraphics{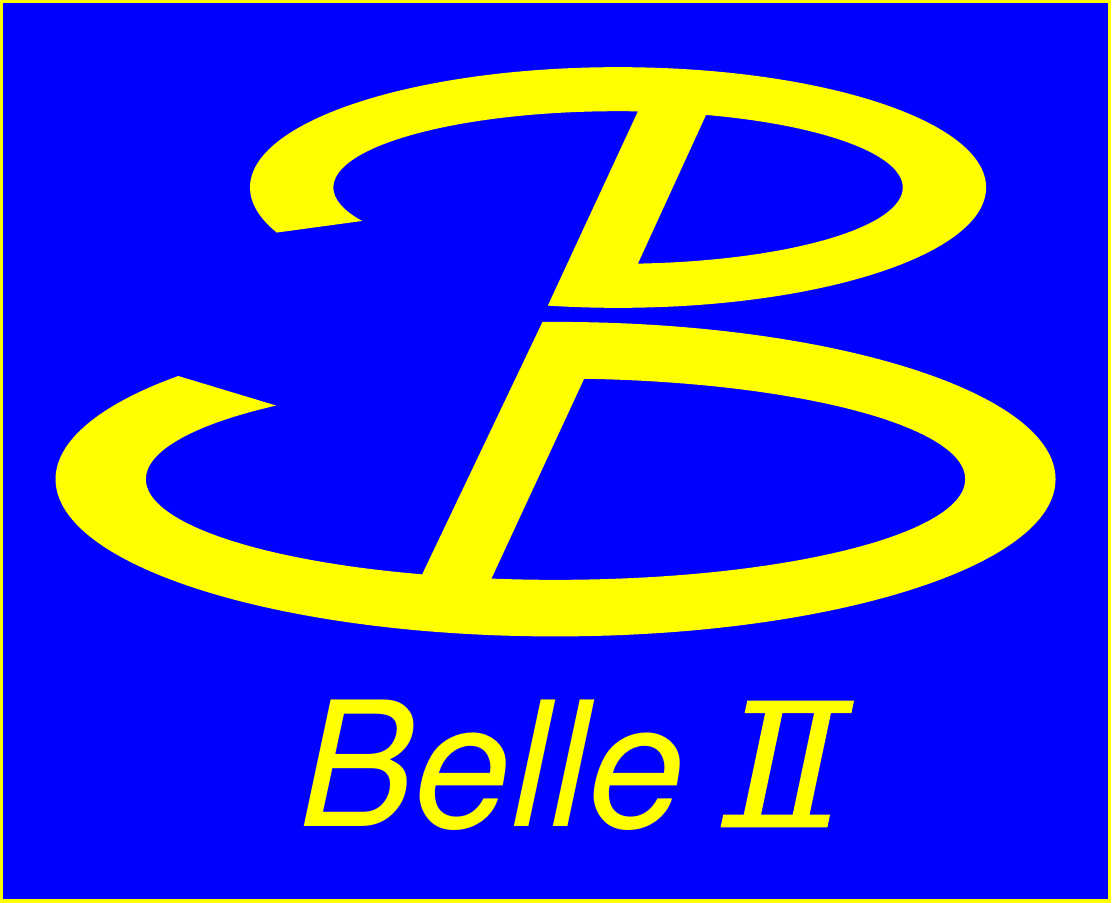}}

\vspace*{-5\baselineskip}
\begin{flushright}
BELLE2-CONF-PH-2021-001

\today
\end{flushright}

\quad\\[0.5cm]

\title {First search for direct \CP-violating asymmetry in $\Bz \to \Kz \piz$ decays at Belle II} 

\newcommand{\instCPPM}{Aix Marseille Universit\'{e}, CNRS/IN2P3, CPPM, 13288 Marseille, France}
\newcommand{\instBeihang}{Beihang University, Beijing 100191, China}
\newcommand{\instBNL}{Brookhaven National Laboratory, Upton, New York 11973, U.S.A.}
\newcommand{\instBINP}{Budker Institute of Nuclear Physics SB RAS, Novosibirsk 630090, Russian Federation}
\newcommand{\instCMU}{Carnegie Mellon University, Pittsburgh, Pennsylvania 15213, U.S.A.}
\newcommand{\instCinvestavIPN}{Centro de Investigacion y de Estudios Avanzados del Instituto Politecnico Nacional, Mexico City 07360, Mexico}
\newcommand{\instPrague}{Faculty of Mathematics and Physics, Charles University, 121 16 Prague, Czech Republic}
\newcommand{\instChiangMai}{Chiang Mai University, Chiang Mai 50202, Thailand}
\newcommand{\instChiba}{Chiba University, Chiba 263-8522, Japan}
\newcommand{\instChonnam}{Chonnam National University, Gwangju 61186, South Korea}
\newcommand{\instConacyt}{Consejo Nacional de Ciencia y Tecnolog\'{\i}a, Mexico City 03940, Mexico}
\newcommand{\instDESY}{Deutsches Elektronen--Synchrotron, 22607 Hamburg, Germany}
\newcommand{\instDuke}{Duke University, Durham, North Carolina 27708, U.S.A.}
\newcommand{\instITAR}{Institute of Theoretical and Applied Research (ITAR), Duy Tan University, Hanoi 100000, Vietnam}
\newcommand{\instRomaENEA}{ENEA Casaccia, I-00123 Roma, Italy}
\newcommand{\instEri}{Earthquake Research Institute, University of Tokyo, Tokyo 113-0032, Japan}
\newcommand{\instJuelich}{Forschungszentrum J\"{u}lich, 52425 J\"{u}lich, Germany}
\newcommand{\instFuJen}{Department of Physics, Fu Jen Catholic University, Taipei 24205, Taiwan}
\newcommand{\instFudan}{Key Laboratory of Nuclear Physics and Ion-beam Application (MOE) and Institute of Modern Physics, Fudan University, Shanghai 200443, China}
\newcommand{\instGoettingen}{II. Physikalisches Institut, Georg-August-Universit\"{a}t G\"{o}ttingen, 37073 G\"{o}ttingen, Germany}
\newcommand{\instGifu}{Gifu University, Gifu 501-1193, Japan}
\newcommand{\instSOKENDAI}{The Graduate University for Advanced Studies (SOKENDAI), Hayama 240-0193, Japan}
\newcommand{\instGyeongsang}{Gyeongsang National University, Jinju 52828, South Korea}
\newcommand{\instHanyang}{Department of Physics and Institute of Natural Sciences, Hanyang University, Seoul 04763, South Korea}
\newcommand{\instKEK}{High Energy Accelerator Research Organization (KEK), Tsukuba 305-0801, Japan}
\newcommand{\instJPARC}{J-PARC Branch, KEK Theory Center, High Energy Accelerator Research Organization (KEK), Tsukuba 305-0801, Japan}
\newcommand{\instHSE}{National Research University Higher School of Economics, Moscow 101000, Russian Federation}
\newcommand{\instIISER}{Indian Institute of Science Education and Research Mohali, SAS Nagar, 140306, India}
\newcommand{\instIITBhubaneswar}{Indian Institute of Technology Bhubaneswar, Satya Nagar 751007, India}
\newcommand{\instIITGuwahati}{Indian Institute of Technology Guwahati, Assam 781039, India}
\newcommand{\instIITHyderabad}{Indian Institute of Technology Hyderabad, Telangana 502285, India}
\newcommand{\instIITMadras}{Indian Institute of Technology Madras, Chennai 600036, India}
\newcommand{\instIndiana}{Indiana University, Bloomington, Indiana 47408, U.S.A.}
\newcommand{\instIHEPRussia}{Institute for High Energy Physics, Protvino 142281, Russian Federation}
\newcommand{\instHEPHYVienna}{Institute of High Energy Physics, Vienna 1050, Austria}
\newcommand{\instHiroshima}{Hiroshima University, Higashi-Hiroshima, Hiroshima 739-8530, Japan}
\newcommand{\instIHEPChina}{Institute of High Energy Physics, Chinese Academy of Sciences, Beijing 100049, China}
\newcommand{\instIPP}{Institute of Particle Physics (Canada), Victoria, British Columbia V8W 2Y2, Canada}
\newcommand{\instIOP}{Institute of Physics, Vietnam Academy of Science and Technology (VAST), Hanoi, Vietnam}
\newcommand{\instIFIC}{Instituto de Fisica Corpuscular, Paterna 46980, Spain}
\newcommand{\instFrascati}{INFN Laboratori Nazionali di Frascati, I-00044 Frascati, Italy}
\newcommand{\instNapoliINFN}{INFN Sezione di Napoli, I-80126 Napoli, Italy}
\newcommand{\instPadovaINFN}{INFN Sezione di Padova, I-35131 Padova, Italy}
\newcommand{\instPerugiaINFN}{INFN Sezione di Perugia, I-06123 Perugia, Italy}
\newcommand{\instPisaINFN}{INFN Sezione di Pisa, I-56127 Pisa, Italy}
\newcommand{\instRomaINFN}{INFN Sezione di Roma, I-00185 Roma, Italy}
\newcommand{\instRomaTreINFN}{INFN Sezione di Roma Tre, I-00146 Roma, Italy}
\newcommand{\instTorinoINFN}{INFN Sezione di Torino, I-10125 Torino, Italy}
\newcommand{\instTriesteINFN}{INFN Sezione di Trieste, I-34127 Trieste, Italy}
\newcommand{\instJAEA}{Advanced Science Research Center, Japan Atomic Energy Agency, Naka 319-1195, Japan}
\newcommand{\instMainz}{Johannes Gutenberg-Universit\"{a}t Mainz, Institut f\"{u}r Kernphysik, D-55099 Mainz, Germany}
\newcommand{\instGiessen}{Justus-Liebig-Universit\"{a}t Gie\ss{}en, 35392 Gie\ss{}en, Germany}
\newcommand{\instKarlsruhe}{Institut f\"{u}r Experimentelle Teilchenphysik, Karlsruher Institut f\"{u}r Technologie, 76131 Karlsruhe, Germany}
\newcommand{\instISU}{Iowa State University, Ames, Iowa 50011, U.S.A.}
\newcommand{\instKitasato}{Kitasato University, Sagamihara 252-0373, Japan}
\newcommand{\instKISTI}{Korea Institute of Science and Technology Information, Daejeon 34141, South Korea}
\newcommand{\instKoreaUnivKU}{Korea University, Seoul 02841, South Korea}
\newcommand{\instKSU}{Kyoto Sangyo University, Kyoto 603-8555, Japan}
\newcommand{\instKyungpook}{Kyungpook National University, Daegu 41566, South Korea}
\newcommand{\instLPI}{P.N. Lebedev Physical Institute of the Russian Academy of Sciences, Moscow 119991, Russian Federation}
\newcommand{\instLNNU}{Liaoning Normal University, Dalian 116029, China}
\newcommand{\instLMU}{Ludwig Maximilians University, 80539 Munich, Germany}
\newcommand{\instLuther}{Luther College, Decorah, Iowa 52101, U.S.A.}
\newcommand{\instMNITJaipur}{Malaviya National Institute of Technology Jaipur, Jaipur 302017, India}
\newcommand{\instMPP}{Max-Planck-Institut f\"{u}r Physik, 80805 M\"{u}nchen, Germany}
\newcommand{\instMPGHLL}{Semiconductor Laboratory of the Max Planck Society, 81739 M\"{u}nchen, Germany}
\newcommand{\instMcGill}{McGill University, Montr\'{e}al, Qu\'{e}bec, H3A 2T8, Canada}
\newcommand{\instMEPhI}{Moscow Physical Engineering Institute, Moscow 115409, Russian Federation}
\newcommand{\instNagoya}{Graduate School of Science, Nagoya University, Nagoya 464-8602, Japan}
\newcommand{\instNagoyaIAR}{Institute for Advanced Research, Nagoya University, Nagoya 464-8602, Japan}
\newcommand{\instNagoyaKMI}{Kobayashi-Maskawa Institute, Nagoya University, Nagoya 464-8602, Japan}
\newcommand{\instNaraWu}{Nara Women's University, Nara 630-8506, Japan}
\newcommand{\instNTUTaiwan}{Department of Physics, National Taiwan University, Taipei 10617, Taiwan}
\newcommand{\instNUUTaiwan}{National United University, Miao Li 36003, Taiwan}
\newcommand{\instKrakow}{H. Niewodniczanski Institute of Nuclear Physics, Krakow 31-342, Poland}
\newcommand{\instNiigata}{Niigata University, Niigata 950-2181, Japan}
\newcommand{\instNSU}{Novosibirsk State University, Novosibirsk 630090, Russian Federation}
\newcommand{\instOkinawa}{Okinawa Institute of Science and Technology, Okinawa 904-0495, Japan}
\newcommand{\instOsakaCity}{Osaka City University, Osaka 558-8585, Japan}
\newcommand{\instRCNP}{Research Center for Nuclear Physics, Osaka University, Osaka 567-0047, Japan}
\newcommand{\instPNNL}{Pacific Northwest National Laboratory, Richland, Washington 99352, U.S.A.}
\newcommand{\instPanjab}{Panjab University, Chandigarh 160014, India}
\newcommand{\instPanjabPAU}{Punjab Agricultural University, Ludhiana 141004, India}
\newcommand{\instRIKENMSL}{Meson Science Laboratory, Cluster for Pioneering Research, RIKEN, Saitama 351-0198, Japan}
\newcommand{\instSeoul}{Seoul National University, Seoul 08826, South Korea}
\newcommand{\instSPU}{Showa Pharmaceutical University, Tokyo 194-8543, Japan}
\newcommand{\instSoochow}{Soochow University, Suzhou 215006, China}
\newcommand{\instSoongsil}{Soongsil University, Seoul 06978, South Korea}
\newcommand{\instLjubljanaJSI}{J. Stefan Institute, 1000 Ljubljana, Slovenia}
\newcommand{\instKyiv}{Taras Shevchenko National Univ. of Kiev, Kiev, Ukraine}
\newcommand{\instTata}{Tata Institute of Fundamental Research, Mumbai 400005, India}
\newcommand{\instTUM}{Department of Physics, Technische Universit\"{a}t M\"{u}nchen, 85748 Garching, Germany}
\newcommand{\instTelAviv}{Tel Aviv University, School of Physics and Astronomy, Tel Aviv, 69978, Israel}
\newcommand{\instToho}{Toho University, Funabashi 274-8510, Japan}
\newcommand{\instTohoku}{Department of Physics, Tohoku University, Sendai 980-8578, Japan}
\newcommand{\instTitech}{Tokyo Institute of Technology, Tokyo 152-8550, Japan}
\newcommand{\instTokyoMetropolitan}{Tokyo Metropolitan University, Tokyo 192-0397, Japan}
\newcommand{\instUAS}{Universidad Autonoma de Sinaloa, Sinaloa 80000, Mexico}
\newcommand{\instNapoliUNIV}{Dipartimento di Scienze Fisiche, Universit\`{a} di Napoli Federico II, I-80126 Napoli, Italy}
\newcommand{\instPadovaUNIV}{Dipartimento di Fisica e Astronomia, Universit\`{a} di Padova, I-35131 Padova, Italy}
\newcommand{\instPerugiaUNIV}{Dipartimento di Fisica, Universit\`{a} di Perugia, I-06123 Perugia, Italy}
\newcommand{\instPisaUNIV}{Dipartimento di Fisica, Universit\`{a} di Pisa, I-56127 Pisa, Italy}
\newcommand{\instRomaTreUNIV}{Dipartimento di Matematica e Fisica, Universit\`{a} di Roma Tre, I-00146 Roma, Italy}
\newcommand{\instTorinoUNIV}{Dipartimento di Fisica, Universit\`{a} di Torino, I-10125 Torino, Italy}
\newcommand{\instTriesteUNIV}{Dipartimento di Fisica, Universit\`{a} di Trieste, I-34127 Trieste, Italy}
\newcommand{\instMontreal}{Universit\'{e} de Montr\'{e}al, Physique des Particules, Montr\'{e}al, Qu\'{e}bec, H3C 3J7, Canada}
\newcommand{\instIJCLab}{Universit\'{e} Paris-Saclay, CNRS/IN2P3, IJCLab, 91405 Orsay, France}
\newcommand{\instIPHC}{Universit\'{e} de Strasbourg, CNRS, IPHC, UMR 7178, 67037 Strasbourg, France}
\newcommand{\instAdelaide}{Department of Physics, University of Adelaide, Adelaide, South Australia 5005, Australia}
\newcommand{\instBonn}{University of Bonn, 53115 Bonn, Germany}
\newcommand{\instUBC}{University of British Columbia, Vancouver, British Columbia, V6T 1Z1, Canada}
\newcommand{\instCincinnati}{University of Cincinnati, Cincinnati, Ohio 45221, U.S.A.}
\newcommand{\instFlorida}{University of Florida, Gainesville, Florida 32611, U.S.A.}
\newcommand{\instHawaii}{University of Hawaii, Honolulu, Hawaii 96822, U.S.A.}
\newcommand{\instHeidelberg}{University of Heidelberg, 68131 Mannheim, Germany}
\newcommand{\instLjubljanaUniLJ}{Faculty of Mathematics and Physics, University of Ljubljana, 1000 Ljubljana, Slovenia}
\newcommand{\instLouisville}{University of Louisville, Louisville, Kentucky 40292, U.S.A.}
\newcommand{\instMalaya}{National Centre for Particle Physics, University Malaya, 50603 Kuala Lumpur, Malaysia}
\newcommand{\instLjubljanaUM}{Faculty of Chemistry and Chemical Engineering, University of Maribor, 2000 Maribor, Slovenia}
\newcommand{\instMelbourne}{School of Physics, University of Melbourne, Victoria 3010, Australia}
\newcommand{\instMississippi}{University of Mississippi, University, Mississippi 38677, U.S.A.}
\newcommand{\instUOM}{University of Miyazaki, Miyazaki 889-2192, Japan}
\newcommand{\instPittsburgh}{University of Pittsburgh, Pittsburgh, Pennsylvania 15260, U.S.A.}
\newcommand{\instUSTC}{University of Science and Technology of China, Hefei 230026, China}
\newcommand{\instSAlabama}{University of South Alabama, Mobile, Alabama 36688, U.S.A.}
\newcommand{\instSCarolina}{University of South Carolina, Columbia, South Carolina 29208, U.S.A.}
\newcommand{\instSydney}{School of Physics, University of Sydney, New South Wales 2006, Australia}
\newcommand{\instUTokyo}{Department of Physics, University of Tokyo, Tokyo 113-0033, Japan}
\newcommand{\instIPMU}{Kavli Institute for the Physics and Mathematics of the Universe (WPI), University of Tokyo, Kashiwa 277-8583, Japan}
\newcommand{\instVictoria}{University of Victoria, Victoria, British Columbia, V8W 3P6, Canada}
\newcommand{\instVPI}{Virginia Polytechnic Institute and State University, Blacksburg, Virginia 24061, U.S.A.}
\newcommand{\instWayneState}{Wayne State University, Detroit, Michigan 48202, U.S.A.}
\newcommand{\instYamagata}{Yamagata University, Yamagata 990-8560, Japan}
\newcommand{\instYerevan}{Alikhanyan National Science Laboratory, Yerevan 0036, Armenia}
\newcommand{\instYonsei}{Yonsei University, Seoul 03722, South Korea}
\newcommand{\instZZU}{Zhengzhou University, Zhengzhou 450001, China}
\affiliation{\instCPPM}
\affiliation{\instBeihang}
\affiliation{\instBNL}
\affiliation{\instBINP}
\affiliation{\instCMU}
\affiliation{\instCinvestavIPN}
\affiliation{\instPrague}
\affiliation{\instChiangMai}
\affiliation{\instChiba}
\affiliation{\instChonnam}
\affiliation{\instConacyt}
\affiliation{\instDESY}
\affiliation{\instDuke}
\affiliation{\instITAR}
\affiliation{\instRomaENEA}
\affiliation{\instEri}
\affiliation{\instJuelich}
\affiliation{\instFuJen}
\affiliation{\instFudan}
\affiliation{\instGoettingen}
\affiliation{\instGifu}
\affiliation{\instSOKENDAI}
\affiliation{\instGyeongsang}
\affiliation{\instHanyang}
\affiliation{\instKEK}
\affiliation{\instJPARC}
\affiliation{\instHSE}
\affiliation{\instIISER}
\affiliation{\instIITBhubaneswar}
\affiliation{\instIITGuwahati}
\affiliation{\instIITHyderabad}
\affiliation{\instIITMadras}
\affiliation{\instIndiana}
\affiliation{\instIHEPRussia}
\affiliation{\instHEPHYVienna}
\affiliation{\instHiroshima}
\affiliation{\instIHEPChina}
\affiliation{\instIPP}
\affiliation{\instIOP}
\affiliation{\instIFIC}
\affiliation{\instFrascati}
\affiliation{\instNapoliINFN}
\affiliation{\instPadovaINFN}
\affiliation{\instPerugiaINFN}
\affiliation{\instPisaINFN}
\affiliation{\instRomaINFN}
\affiliation{\instRomaTreINFN}
\affiliation{\instTorinoINFN}
\affiliation{\instTriesteINFN}
\affiliation{\instJAEA}
\affiliation{\instMainz}
\affiliation{\instGiessen}
\affiliation{\instKarlsruhe}
\affiliation{\instISU}
\affiliation{\instKitasato}
\affiliation{\instKISTI}
\affiliation{\instKoreaUnivKU}
\affiliation{\instKSU}
\affiliation{\instKyungpook}
\affiliation{\instLPI}
\affiliation{\instLNNU}
\affiliation{\instLMU}
\affiliation{\instLuther}
\affiliation{\instMNITJaipur}
\affiliation{\instMPP}
\affiliation{\instMPGHLL}
\affiliation{\instMcGill}
\affiliation{\instMEPhI}
\affiliation{\instNagoya}
\affiliation{\instNagoyaIAR}
\affiliation{\instNagoyaKMI}
\affiliation{\instNaraWu}
\affiliation{\instNTUTaiwan}
\affiliation{\instNUUTaiwan}
\affiliation{\instKrakow}
\affiliation{\instNiigata}
\affiliation{\instNSU}
\affiliation{\instOkinawa}
\affiliation{\instOsakaCity}
\affiliation{\instRCNP}
\affiliation{\instPNNL}
\affiliation{\instPanjab}
\affiliation{\instPanjabPAU}
\affiliation{\instRIKENMSL}
\affiliation{\instSeoul}
\affiliation{\instSPU}
\affiliation{\instSoochow}
\affiliation{\instSoongsil}
\affiliation{\instLjubljanaJSI}
\affiliation{\instKyiv}
\affiliation{\instTata}
\affiliation{\instTUM}
\affiliation{\instTelAviv}
\affiliation{\instToho}
\affiliation{\instTohoku}
\affiliation{\instTitech}
\affiliation{\instTokyoMetropolitan}
\affiliation{\instUAS}
\affiliation{\instNapoliUNIV}
\affiliation{\instPadovaUNIV}
\affiliation{\instPerugiaUNIV}
\affiliation{\instPisaUNIV}
\affiliation{\instRomaTreUNIV}
\affiliation{\instTorinoUNIV}
\affiliation{\instTriesteUNIV}
\affiliation{\instMontreal}
\affiliation{\instIJCLab}
\affiliation{\instIPHC}
\affiliation{\instAdelaide}
\affiliation{\instBonn}
\affiliation{\instUBC}
\affiliation{\instCincinnati}
\affiliation{\instFlorida}
\affiliation{\instHawaii}
\affiliation{\instHeidelberg}
\affiliation{\instLjubljanaUniLJ}
\affiliation{\instLouisville}
\affiliation{\instMalaya}
\affiliation{\instLjubljanaUM}
\affiliation{\instMelbourne}
\affiliation{\instMississippi}
\affiliation{\instUOM}
\affiliation{\instPittsburgh}
\affiliation{\instUSTC}
\affiliation{\instSAlabama}
\affiliation{\instSCarolina}
\affiliation{\instSydney}
\affiliation{\instUTokyo}
\affiliation{\instIPMU}
\affiliation{\instVictoria}
\affiliation{\instVPI}
\affiliation{\instWayneState}
\affiliation{\instYamagata}
\affiliation{\instYerevan}
\affiliation{\instYonsei}
\affiliation{\instZZU}
  \author{F.~Abudin{\'e}n}\affiliation{\instTriesteINFN} 
  \author{I.~Adachi}\affiliation{\instKEK}\affiliation{\instSOKENDAI} 
  \author{R.~Adak}\affiliation{\instFudan} 
  \author{K.~Adamczyk}\affiliation{\instKrakow} 
  \author{P.~Ahlburg}\affiliation{\instBonn} 
  \author{J.~K.~Ahn}\affiliation{\instKoreaUnivKU} 
  \author{H.~Aihara}\affiliation{\instUTokyo} 
  \author{N.~Akopov}\affiliation{\instYerevan} 
  \author{A.~Aloisio}\affiliation{\instNapoliUNIV}\affiliation{\instNapoliINFN} 
  \author{F.~Ameli}\affiliation{\instRomaINFN} 
% ...

\collaboration{Belle II Collaboration}

\begin{abstract}
We report on the first measurement of the direct \CP-violating asymmetry ($\mathcal{A}$) in the charmless decay $B^0 \to K^0\pi^0$ at Belle~II and an updated measurement of its branching fraction~($\mathcal{B}$). We use a sample of electron-positron collisions collected in 2019 and 2020 at the $\Upsilon(4S)$ resonance and  corresponding to $62.8$\,fb$^{-1}$ of integrated luminosity. 
We reconstruct and select about $50$
$\Bz \to \KS \pi^0$ candidates, and we measure $\akzpiz = -0.40_{-0.44}^{+0.46} (\rm stat) \pm 0.04 (\rm syst)$ and $\mathcal{B}(\Bz \to \Kz \piz) = [8.5_{-1.6}^{+1.7} (\rm stat)  \pm 1.2  (\rm syst)] \times 10^{-6}$.
This is the first measurement of \CP violation in $B^0 \to K^0\pi^0$ decays reported by Belle~II. 
The results agree with previous determinations and show a detector performance comparable with the best Belle results.
 
\end{abstract}


\maketitle

{\renewcommand{\thefootnote}{\fnsymbol{footnote}}}
\setcounter{footnote}{0}




\section{Introduction and motivation}

The study of charmless $B$ decays is a keystone of the worldwide flavor program. Processes mediated by $b\to sq\bar{q}$ 
transitions probe contributions of non-standard-model dynamics in loop decay-amplitudes.  However, reliable extraction of weak phases and unambiguous interpretation of measurements involving these amplitudes is spoiled by large hadronic uncertainties, which are rarely tractable in perturbative calculations. Appropriately chosen combinations of measurements from decay modes related by flavor symmetries are used to significantly reduce the impact of such unknowns. An especially  fruitful approach consists in combining measurements from decays related by isospin symmetries. This approach has been proposed to address the so-called $K\pi$ {\it puzzle}, a long-standing anomaly associated with the significant difference between direct \CP-violating asymmetries observed in $B^0 \to K^+\pi^-$ and $B^+ \to K^+\pi^0$ decays~\cite{PhysRevD.59.113002}. The asymmetries are expected to be equal at the leading order in the electroweak perturbation theory, as the two decays differ only by the \emph{spectator} quark. The isospin sum rule
\begin{equation}
\begin{aligned}
I_{K\pi} = \mathcal{A}_{K^+\pi^-} + \mathcal{A}_{K^0\pi^+}\frac{\mathcal{B}(K^0\pi^+)}{\mathcal{B}(K^+\pi^-)}\frac{\tau_{B^0}}{\tau_{B^+}} - 2\mathcal{A}_{K^+\pi^0}\frac{\mathcal{B}(K^+\pi^0)}{\mathcal{B}(K^+\pi^-)}\frac{\tau_{B^0}}{\tau_{B^+}} - 2\mathcal{A}_{K^0\pi^0}\frac{\mathcal{B}(K^0\pi^0)}{\mathcal{B}(K^+\pi^-)}
\label{eq:SumRule}
\end{aligned}
\end{equation}
properly accounts for subleading amplitudes by combining the branching fractions ($\mathcal{B}$) and direct \CP-violating asymmetries ($\mathcal{A}$) of $B$ decays to all four final states $\Kp \pim$, $\Kz\pip$, $\Kp\piz$ and $\Kz \piz$, and the lifetime ($\tau$) ratio between \Bp and \Bz. This rule offers a stringent null test of the standard model (SM), which 
predicts $I_{K\pi} = 0$ in the limit of isospin symmetry and no electroweak penguin (EWP) contributions, and with an uncertainty much below $1\%$ when including SM EWP amplitudes~\cite{Gronau:2005kz,Browder:2008em,Gershon:2006mt}. Belle  II  has the unique  capability  of  studying  jointly,  and within a consistent experimental environment, all relevant final states.

The Belle~II experiment, complete with its vertex detector, started colliding beam operations in March 2019 and is currently ongoing.  The sample of electron-positron collisions used in this work corresponds to an integrated luminosity of $62.8\,\si{fb^{-1}}$
and was collected at the $\Upsilon(4{\rm S})$~resonance.
This \mbox{document} reports the first Belle~II measurement of the direct \CP-violating asymmetry in $B^0 \to K^0 \pi^0$  decays, which requires flavor tagging of the pair-produced neutral $B$ partner, and an updated measurement of its branching fraction that supersedes the previous Belle II result in Ref.~\cite{belleiicollaboration2020measurements}.

All analysis procedures are first developed and finalized in simulated data. 
For the branching fraction measurement, we test the analysis on the data subset used in~\cite{belleiicollaboration2020measurements}, corresponding to 55\% of the total sample, prior to the application to the full data set.
A signal selection is applied to suppress the major sources of backgrounds, building on previous work~\cite{belleiicollaboration2020measurements}. An initial fit then determines the sample composition in terms of signal; background from $e^+e^- \to q\bar{q}$ continuum events, where $q$ indicates any quark of the first or second family ($u$, $d$, $s$, and $c$); and  background from non-signal $B$ decays. The fit uses


\begin{itemize}
    \item the energy difference $\Delta E \equiv E^{*}_{B} - \sqrt{s}/2$ between the total energy of the reconstructed $B$ candidate and half of the collision energy, both in the $\Upsilon(4S)$ frame, which discriminates misreconstructed from properly reconstructed $B$ decays.
 \item the modified beam-energy-constrained mass, 
    \begin{equation}
    \mbc' \equiv \sqrt{s/4 - \left( \vec{p}^*(\pi^+\pi^-)_{K^{0}_{S}} + \frac{\vec{p}^*(\gamma\gamma)_{\pi^0}}{p^*(\gamma\gamma)_{\pi^0}} \times \sqrt{(\sqrt{s}/2 - E^*(\pi^+\pi^-)_{K_S^0})^2 - m(\gamma\gamma)_{\pi^0}^2} \right)^2 }
    \label{eq:ModMbc}
    \end{equation}
    where $\vec{p}^*$ and $E^*$ are the three-momentum and the energy of \KS or \piz candidates in the CM frame, respectively. This variable is a modification of the standard beam-energy-constrained mass,  $M_{\rm bc} \equiv \sqrt{s/(4c^4) - (p^{*}_B/c)^2}$, 
    where the $B$ energy is replaced by half of the center-of-mass collision energy (which is more precisely known).
    The variable $\mbc'$ compensates for the correlation between \mbc and \de introduced by the tails associated with the presence of a final-state \piz, and discriminates $B$ meson decays from other backgrounds as the standard $\mbc$ does.
\end{itemize}
The analysis focuses only on \de and $\mbc'$. 
For convenience, in the following we  refer to the modified mass with the symbol \mbc only.

\section{The Belle~II detector}
Belle~II is a $4\pi$ particle-physics spectrometer~\cite{Kou:2018nap, Abe:2010sj}, designed to reconstruct the products of electron-positron collisions produced by the SuperKEKB asymmetric-energy collider~\cite{Akai:2018mbz}, located at the KEK laboratory in Tsukuba, Japan. Belle~II comprises several subdetectors arranged around the interaction space-point in a cylindrical geometry. The innermost subdetector is the vertex detector, which uses position-sensitive silicon layers to sample the trajectories of charged particles (tracks) in the vicinity of the interaction region to extrapolate the decay positions of their long-lived parent particles. The vertex detector includes two inner layers of silicon pixel sensors and four outer layers of silicon microstrip sensors. The second pixel layer is currently incomplete and covers only a small portion of azimuthal angle. Charged-particle momenta and charges are measured by a large-radius, helium-ethane, small-cell central drift chamber, which also offers charged-particle-identification information through a measurement of particles' energy-loss by specific ionization. A Cherenkov-light angle and time-of-propagation detector surrounding the chamber provides charged-particle identification in the central detector volume, supplemented by proximity-focusing, aerogel, ring-imaging Cherenkov detectors in the forward regions. A CsI(Tl)-crystal electromagnetic calorimeter allows for energy measurements of electrons and photons.  A solenoid surrounding the calorimeter generates a uniform axial 1.5\,T magnetic field filling its inner volume. Layers of plastic scintillator and resistive-plate chambers, interspersed between the magnetic flux-return iron plates in the barrel, allow for identification of $K^0_{\rm L}$ and muons.
The subdetectors most relevant for this work are the silicon vertex detector, the tracking drift chamber, the particle-identification detectors, and the electromagnetic calorimeter.

\section{Selection and reconstruction}
\label{sec:selection}

\subsection{Simulated and experimental data}
We use generic simulated data to optimize the event selection and compare the distributions observed in experimental data with expectations. We use signal-only simulated data to model relevant signal features for fits and determine selection efficiencies. 
Generic simulation consists of Monte Carlo samples that include $B^0\overline{B}^0$, $B^+B^-$, $u\bar{u}$, $d\bar{d}$, $s\bar{s}$,  and $c\bar{c}$ processes in realistic proportions and corresponding in size to five times the $\Upsilon$(4S) data. In addition, one million $\Bz \to \KS \piz$ signal events are generated.
As for experimental data, we use all 2019--2020  $\Upsilon$(4S) good-quality data collected up to July 1, 2020 and corresponding to an integrated luminosity of $62.8\,\si{fb^{-1}}$. All events are required to satisfy initial loose data-skim selection criteria, based on total energy and charged-particle multiplicity in the event, targeted at reducing sample sizes to a manageable level with negligible impact on signal efficiency. All data are processed using the Belle~II analysis software~\cite{Kuhr:2018lps}.

\subsection{Reconstruction and baseline selection}
We form final-state particle candidates by applying loose baseline selection criteria and then combine candidates in kinematic fits consistent with the topologies of  decays to reconstruct intermediate states and $B$ candidates. 
We reconstruct neutral-pion candidates by combining pairs of photons with energies greater than about $20 \mev$.
We restrict the diphoton mass and excluding extreme helicity-angle values to suppress combinatorial background from collinear soft photons. The mass of the $\pi^0$ candidates is constrained to its known value~\cite{PDG} in subsequent kinematic fits. 
For $K_{\rm S}^0$ reconstruction, we use pairs of oppositely charged particles that originate from a common space-point and have dipion mass consistent with a $K_{\rm S}^0$. To reduce combinatorial background, we apply additional requirements, dependent on $K_{\rm S}^0$~momentum, on the distance between trajectories of the two charged-pion candidates, the $K^0_{\rm S}$~flight distance, and the angle between the pion-pair momentum and the direction of the $K^0_{\rm S}$~flight.
The resulting $\pi^0$ and $\PKzS$ candidates are combined in a simultaneous kinematic fit of the whole decay chain with a constraint on the position of the interaction region to form our target signal channel. 

\subsection{Continuum suppression}
The main challenge in reconstructing  charmless $B$ decay signals is the large contamination from continuum background. We use a binary boosted decision-tree classifier that combines nonlinearly 39 variables known to provide statistical discrimination between $B$-meson signals and continuum background and to be loosely correlated with $\Delta E$ and $M_{\rm bc}$. The variables are quantities associated to event topology (global and signal-only angular configurations).
We train the classifier to identify statistically significant signal and background features using unbiased simulated samples. 

\begin{figure}[h!]
 \centering
    \centering
    \subfigure{\includegraphics[width=0.49\textwidth]{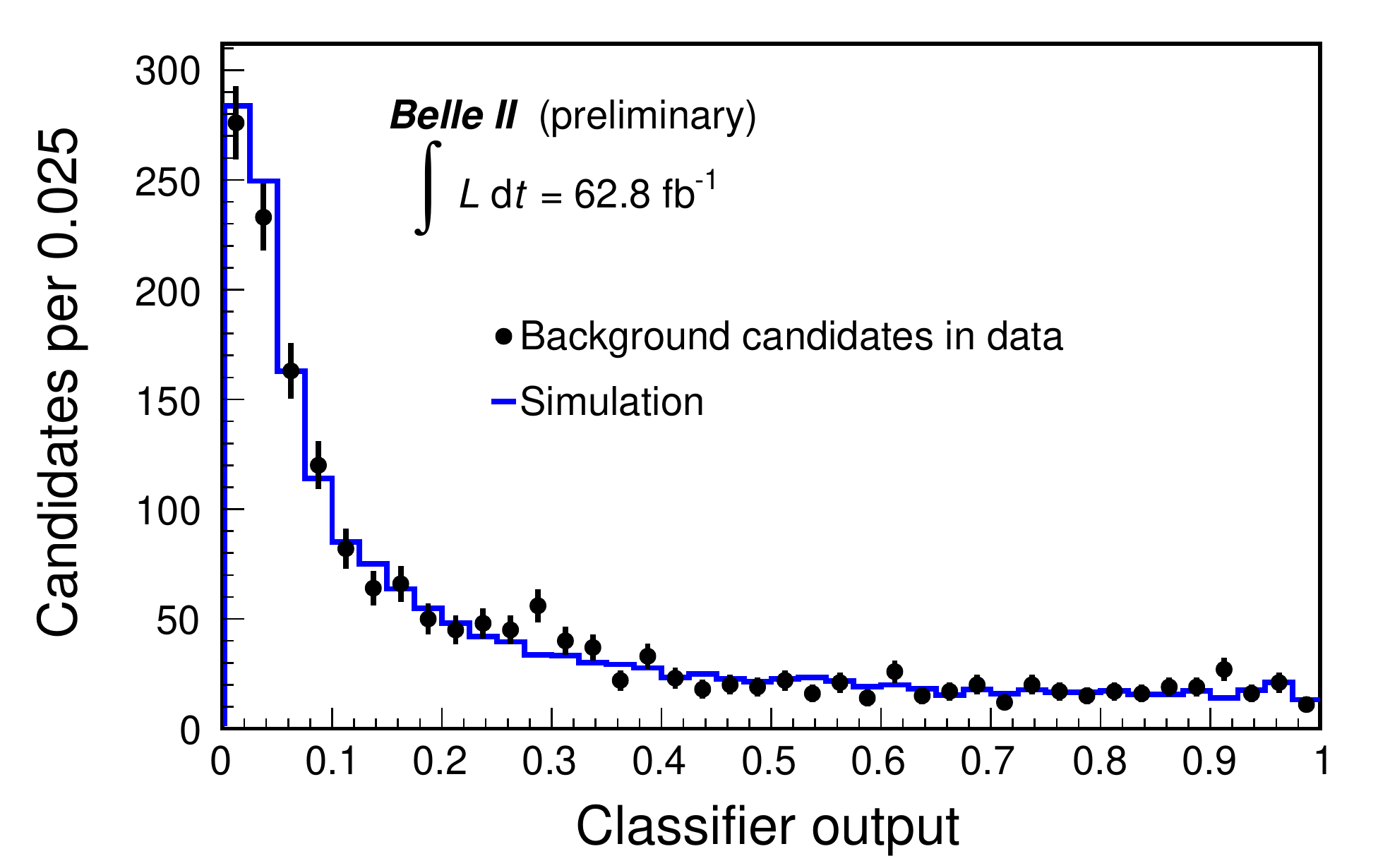}} \hfill
    \subfigure{\includegraphics[width=0.49\textwidth]{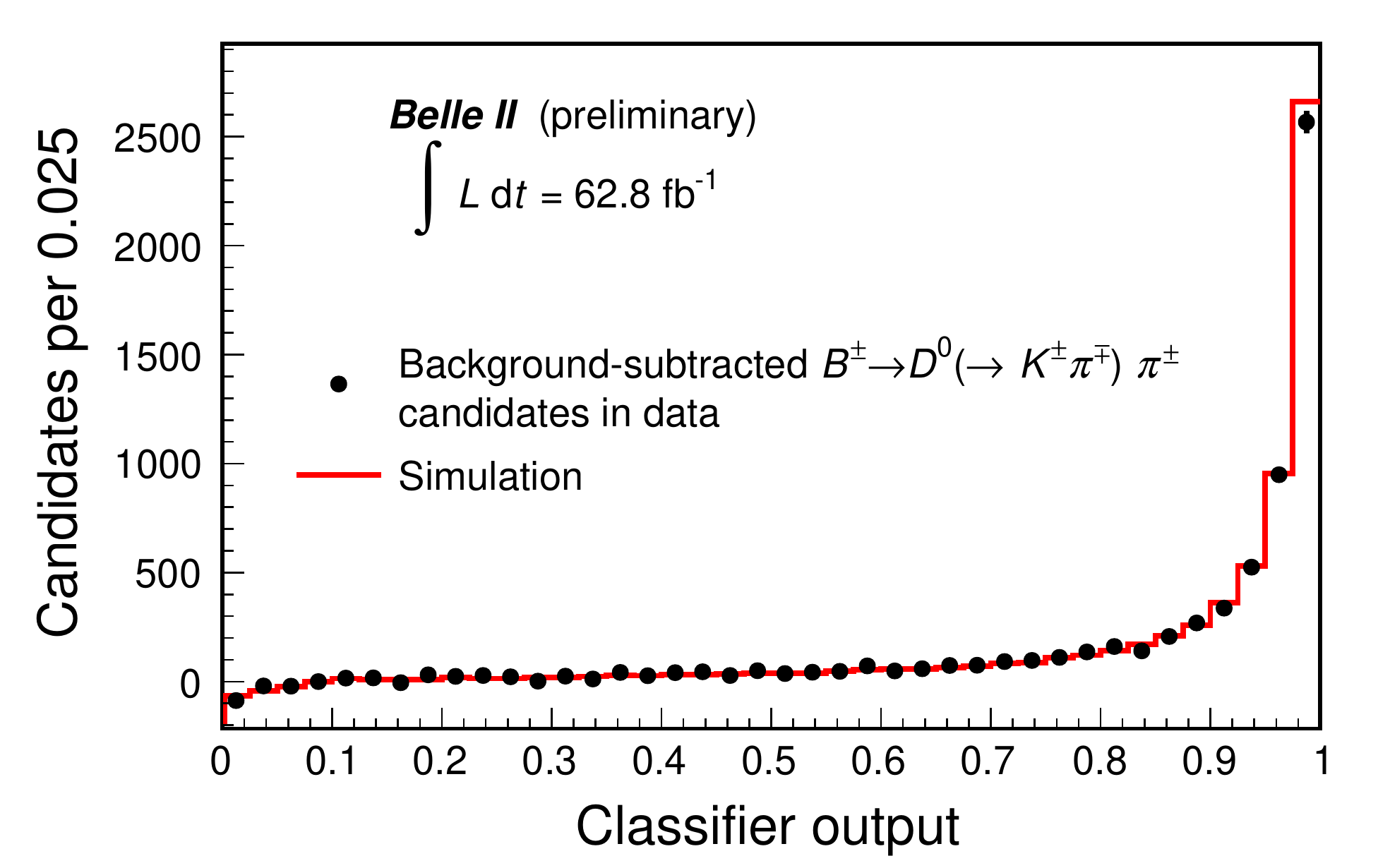}} \\
 \caption{Data-simulation comparison of the output of the boosted decision-tree classifier on (left)~side-band and (right)~side-band-subtracted $\PBplus\to\APD^{0}(\to \PKp\Pgpm)\,\Pgpp$~candidates in the signal region.}
 \label{fig:outputData_Kpi}
\end{figure}

We validate the input and output distributions of the classifier by comparing control sample data with simulation.
Fig.~\ref{fig:outputData_Kpi} shows the distribution of the output for \mbox{$\PBplus\to\APD^{0}(\to \PKp\Pgpm)\,\Pgpp$} candidates reconstructed in data and simulation. No inconsistency is observed. 

\section{Optimization of the signal selection}
\label{sec:opt}
We vary the requirements on photon energy (independently for end-cap and barrel photons), its helicity angle, the diphoton mass range, dipion mass range, and the continuum-suppression output, to maximize ${\rm S}/\sqrt{{\rm S}+{\rm B}}$ using simulated data, where ${\rm S}$ and ${\rm B}$ are signal and background yields, respectively. 
The optimal requirements select energy larger than about $220$ ($80$) \mev for endcap (barrel) photons; absolute cosine values of the photon helicity angles smaller than $0.953$, and diphoton and dipion mass ranges $119$--$150$ \mev and $482$--$513$ \mev, respectively.
The optimal continuum-suppression requirement  suppresses more than $98\%$ of continuum background while retaining about $50\%$ signal efficiency.

\section{Determination of signal yields}
\label{sec:yields}
More than one candidate per event populates the selected sample, with average multiplicities of approximately 1.01.
We restrict to one candidate per event by selecting the $\pi^0$ candidate with the highest $p$-value of the mass-constrained diphoton fit. 
If multiple candidates still remain, the $\KS$ with the best vertex fit $p$-value is used.
Signal yields are determined with maximum likelihood fits of the unbinned two-dimensional \mbc-\de distributions of candidates within the analysis range $M_{\rm bc} > 5.24$\,GeV/$c^2$ and $-0.3 < \Delta E < 0.3$ GeV. 
Plots in this document are projected into the signal region, defined as $M_{\rm bc} > 5.27$\,GeV/$c^2$ for the \de projection or $-0.16 < \Delta E < 0.08$ GeV for the \mbc projection, to clearly display the signal.

\begin{figure}[htb]
 \centering
 \includegraphics[width=0.48\textwidth]{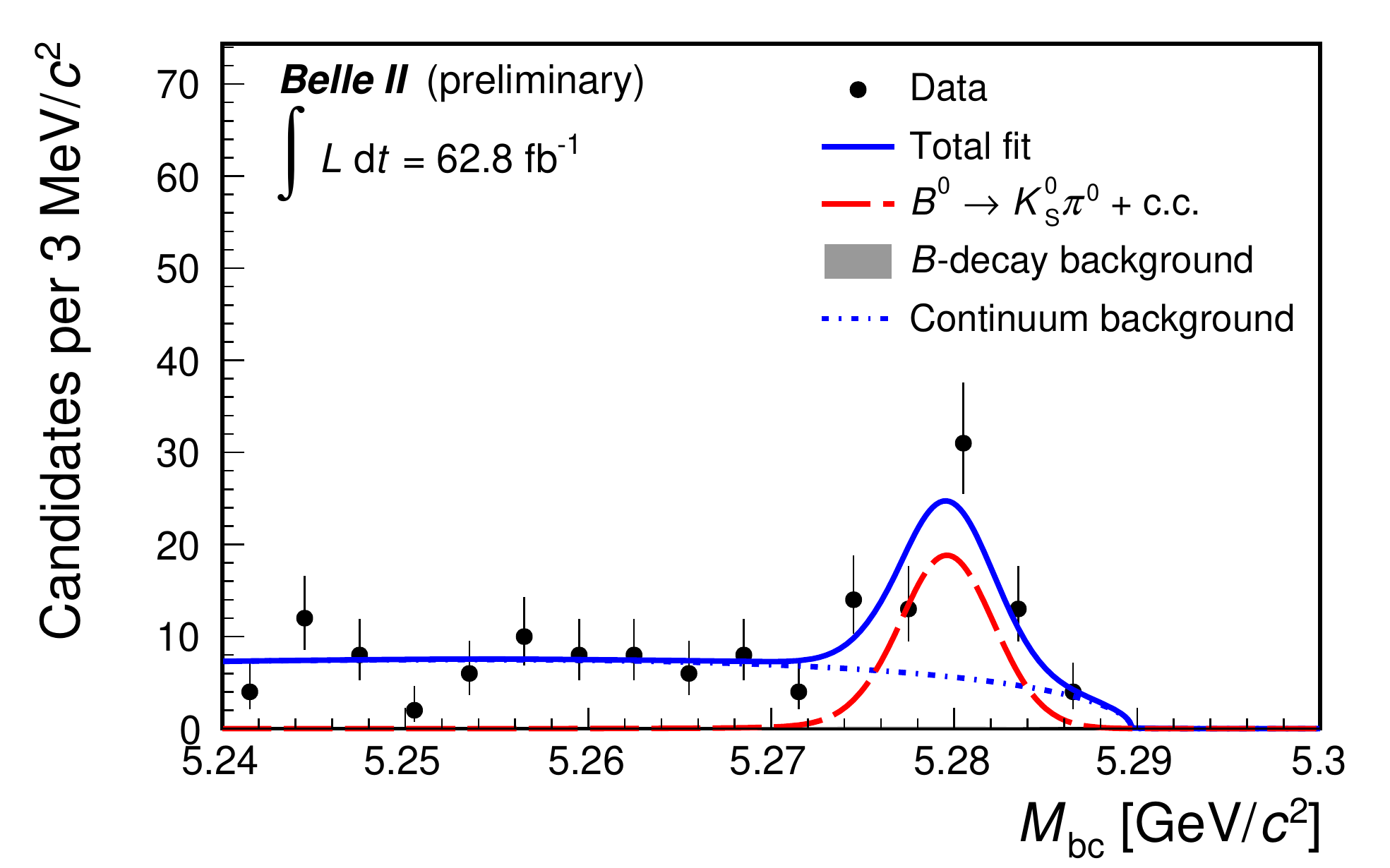}
 \includegraphics[width=0.48\textwidth]{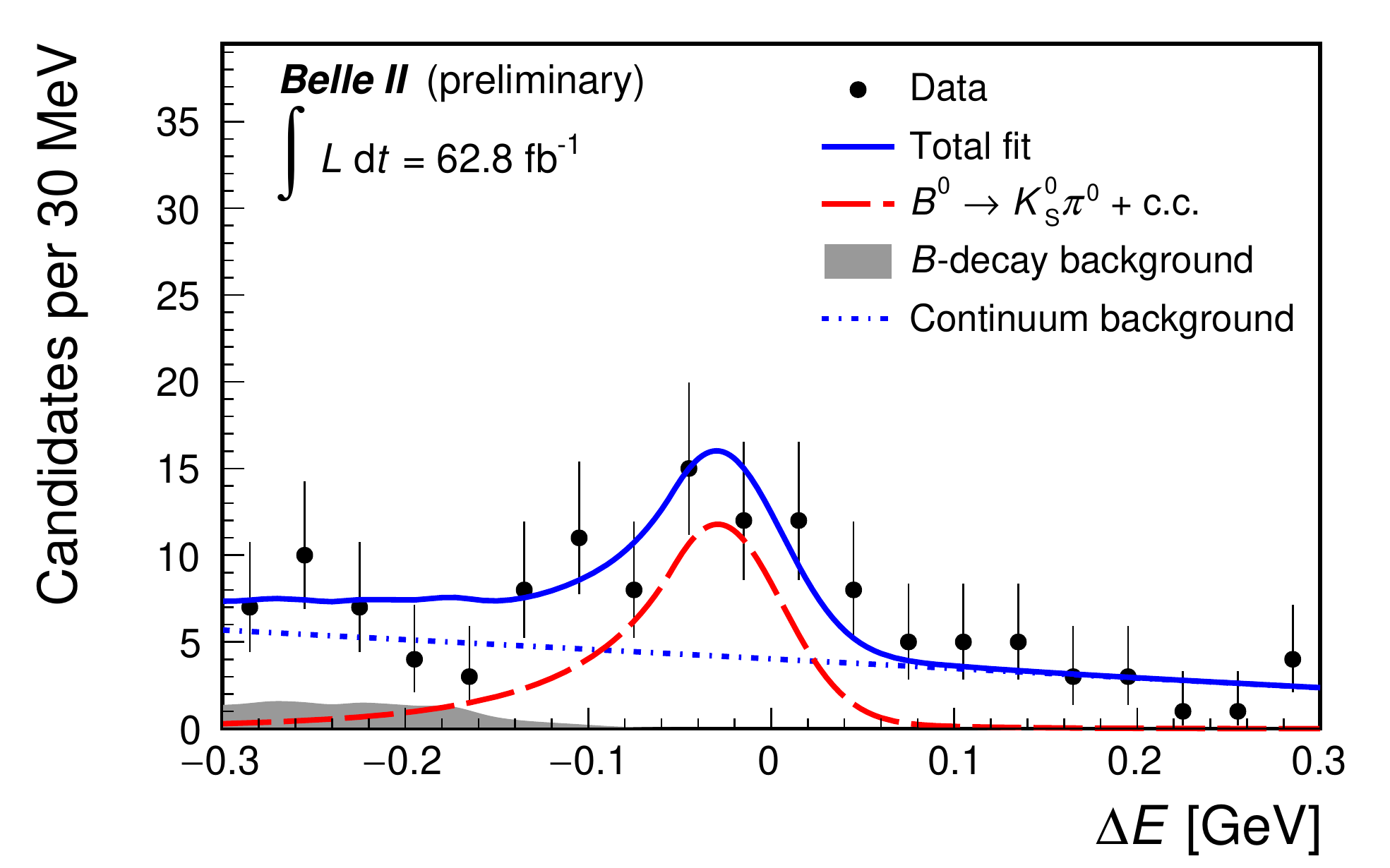}
 \caption{Distribution of $\mbc$ and $\Delta E$ for $B^0 \to \PKzS\pi^0$  candidates reconstructed in 
 2019--2020 Belle~II data selected through the baseline criteria with an optimized continuum-suppression, and projected onto the signal region (left panel: $-0.16 < \Delta E < 0.08$ GeV, right panel: $M_{\rm bc} > 5.27$\,GeV/$c^2$). The projection of the unbinned maximum likelihood fit is overlaid.} 
 \label{fig:KSpi0}
\end{figure}

Fit models are determined from Monte Carlo simulation, with the only additional flexibility of a \de peak position shift determined from $\Bp \to \Kp \piz$ data to account for a negative O($10$) MeV bias in the \piz energy calibration. 
For the signal component, we use the sum of a single or a double Gaussian and a Crystal Ball model~\cite{Skwarnicki:1986xj} for the \mbc and \de distributions, respectively.
For the continuum background, the \mbc distribution is modelled by an Argus function~\cite{Albrecht:1990am}, and that of \de is modelled by a linear function.
A small fraction of non-signal $B$ decays survive the selection, dominated by the decays $\Bp \to \rho^{+} \Kz$ and $\Bp \to K^{*}(892)^{+} \piz$ and totaling $12.7 \pm 1.1$ decays as estimated from simulation. These are lumped together into a single ``$B$-decay background'' component peaking under the signal in \mbc and shifted toward lower \de values, which is modelled by a two-dimensional kernel distribution. The yield of this component is Gaussian-constrained to its estimated value in the fit to data. 
The \mbc-\de distributions of $B^0 \to \KS \piz$ data are shown in Fig.~\ref{fig:KSpi0}, with the fit curve overlaid.

\subsection{Efficiencies and determination of branching fractions}
The signal efficiency determined from simulation is around $16\%$ after all selections.
For those efficiency factors for which simulation may not accurately model data, we perform dedicated checks on control samples of data and assess systematic uncertainties in Sec.~\ref{sec:syst}.

We determine the branching fraction as 

\begin{equation}
\begin{aligned}
\mathcal{B} &= \frac{N_{sig}}{\varepsilon \cdot \mathcal{B}_{s} \cdot 2 \cdot N_{B\bar{B}}}\\
\end{aligned}
\label{eq:BF}
\end{equation}
where $N_{sig}$ is the signal yield obtained from the fit, $\varepsilon$ is the signal efficiency, and $N_{\PB\APB}$ is the number of produced ${\PB\APB}$~pairs, corresponding to $33.9$~million $\PBzero\APBzero$ pairs. We obtain the number of ${\Bz\Bzb}$~pairs from the measured integrated luminosity, the \mbox{$\Pep\Pem\to\Upsilon(4{\rm S})$}~cross section~$(1.110 \pm 0.008)\,$nb~\cite{Bevan:2014iga}~(assuming that the $\Upsilon(4{\rm S})$ decays exclusively to ${\PB\APB}$~pairs), and the \mbox{$\Upsilon(4{\rm S})\to\PBzero\APBzero$}~branching fraction \mbox{$f^{00} = 0.487\pm 0.010\pm 0.008$}~\cite{Aubert:2005bq}. 
The symbol $\mathcal{B}_{s}$ accounts for the $50\%$ probability to have a \KS meson from a \Kz.

\begin{table}[!ht]
    \centering
    \caption{Summary of signal efficiency $\varepsilon$, probability of a $K^0$ decays into a \KS meson $\mathcal{B}_{\rm s} = f(K^0 \to K^0_S)$, signal yield in 2019-2020 Belle~II data, and the resulting branching fraction. Only the statistical contributions to the uncertainties are given here.} 
\begin{tabular}{l   r  r  r  r }
\hline\hline
Decay & \multicolumn{1}{c}{$\varepsilon\,[\%]$} & \multicolumn{1}{c}{$\mathcal{B}_{\rm s}\,[\%]$} & \multicolumn{1}{c}{Yield} & \multicolumn{1}{c}{$\mathcal{B}\,[10^{-6}]$} \\\hline
  $B^0 \to K^0\pi^0$        & $15.6\quad$ &  $50\;\;$   & $45^{+9}_{-8}\;\;\;$         & 		$8.5_{-1.6}^{+1.7}\;$ \\
   \hline\hline
\end{tabular}
    \label{tab:EffYieldBFSummary}
\end{table}

\section{Decay-Time-Integrated Direct \akzpiz}
\label{sec:acp}
 
\begin{figure}[h!]
\centering
    \subfigure{\includegraphics[width=0.48\textwidth]{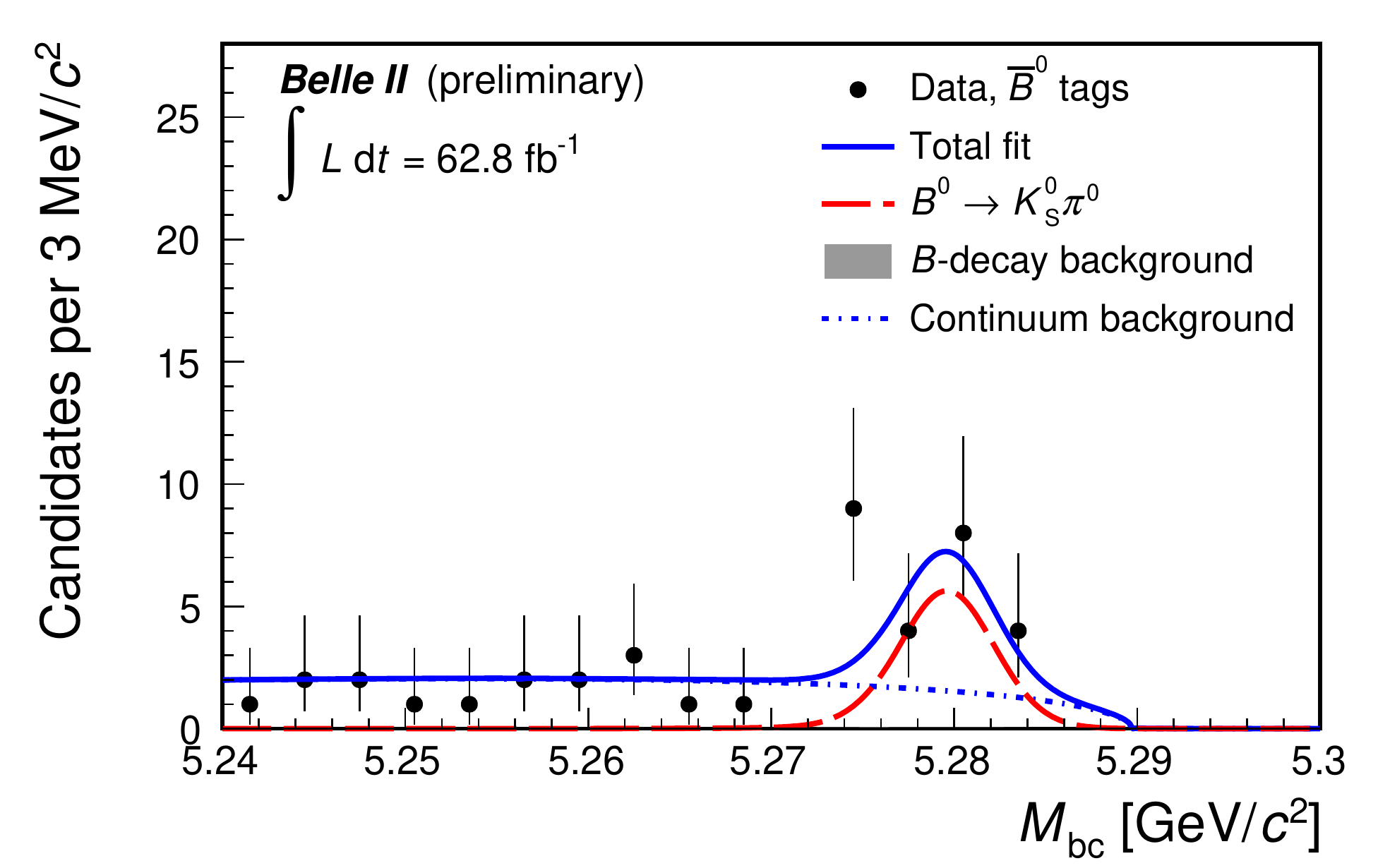}}
    \subfigure{\includegraphics[width=0.48\textwidth]{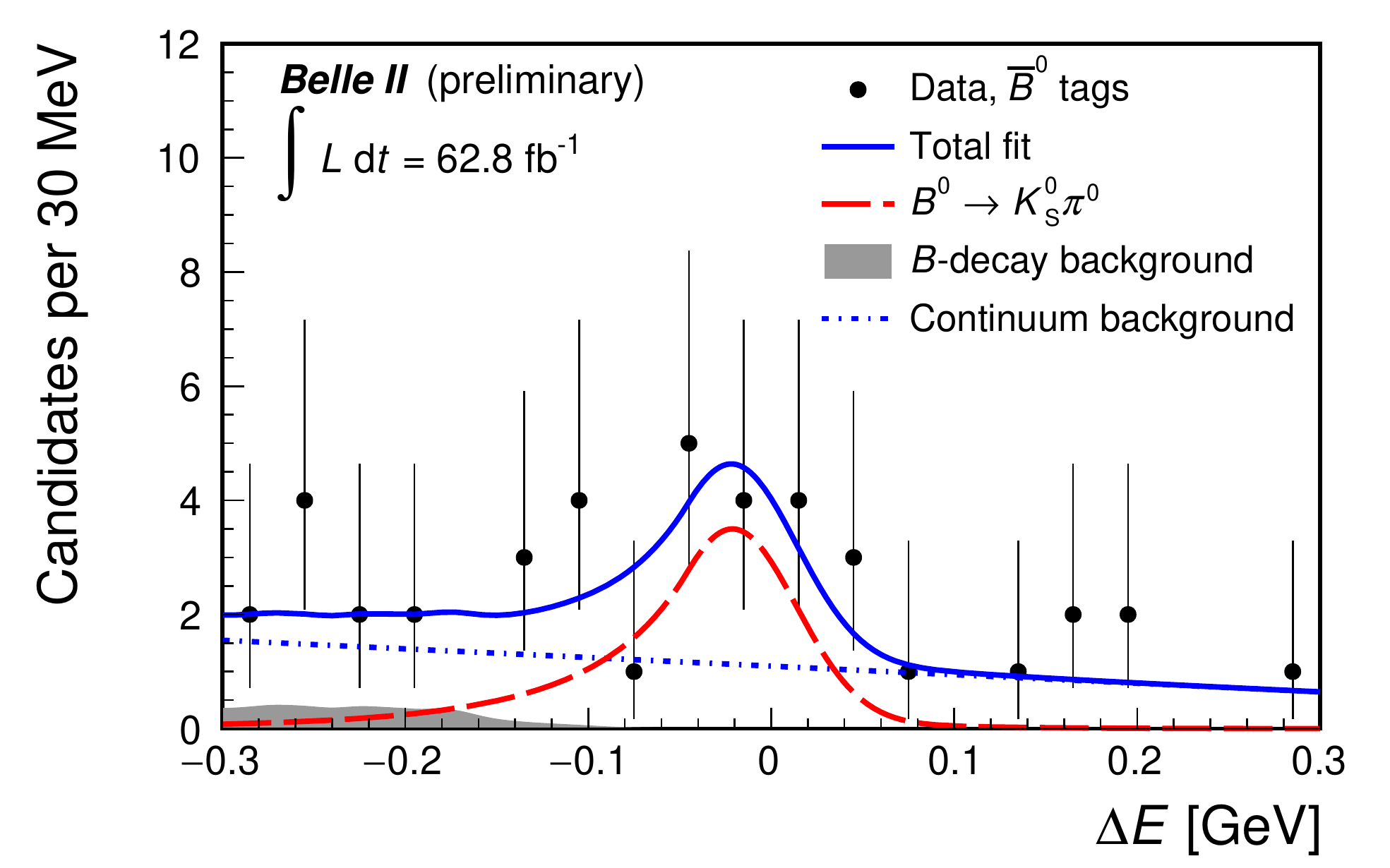}} \hfill
    \subfigure{\includegraphics[width=0.48\textwidth]{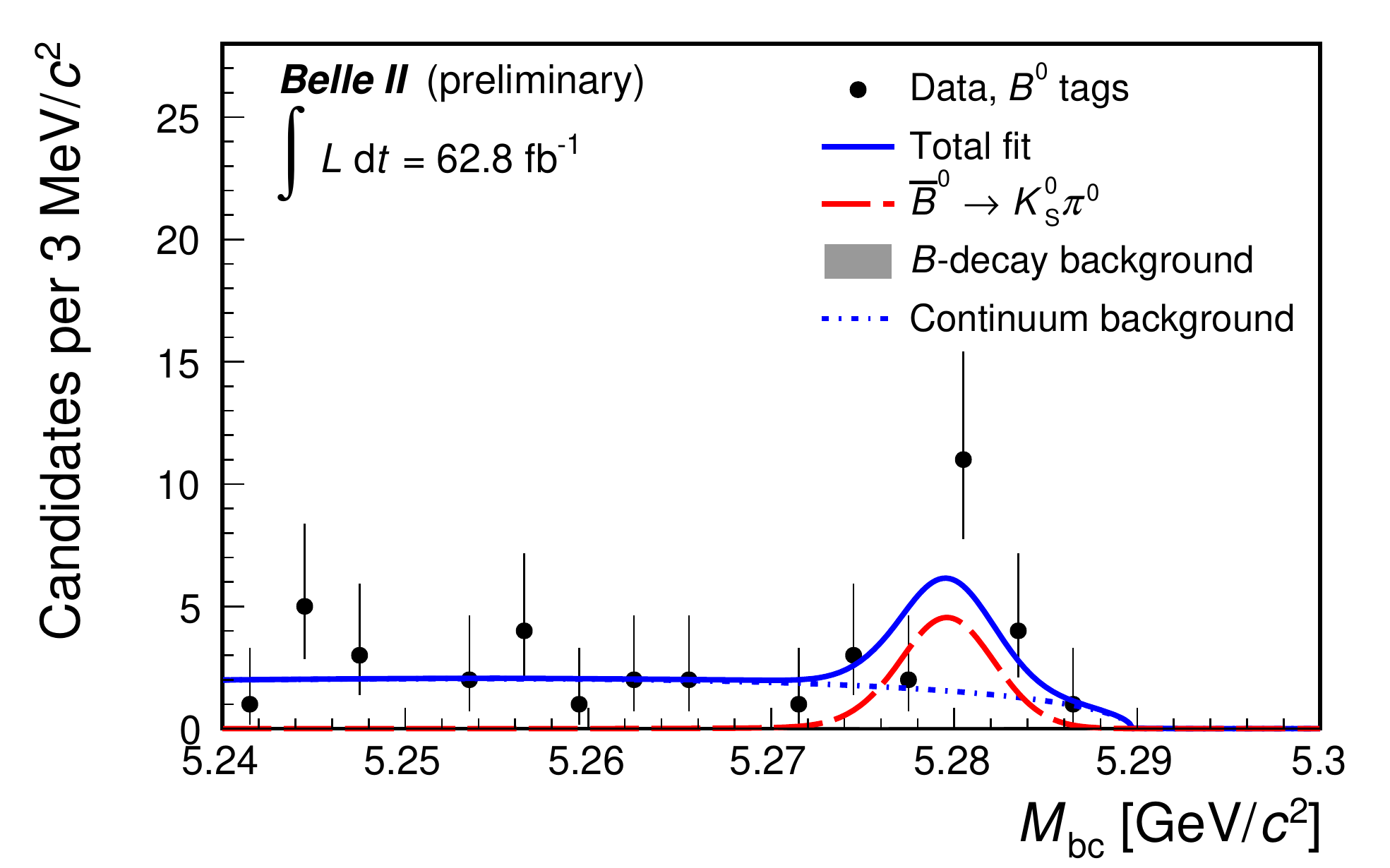}}
    \subfigure{\includegraphics[width=0.48\textwidth]{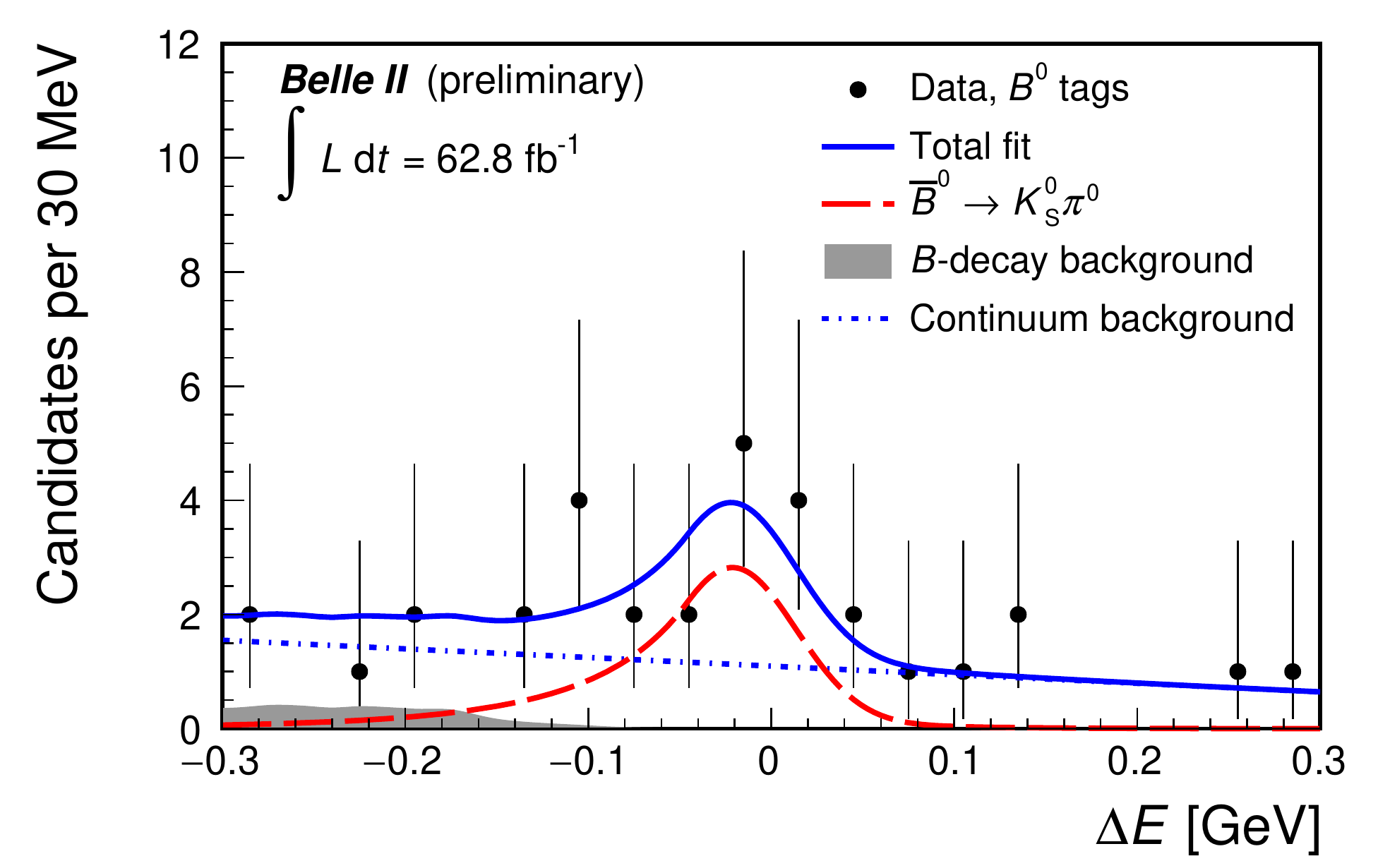}} \\
\caption{Flavor-specific (\mbc, \de) projections on 2019-2020 Belle~II data. The top panel shows candidates where $B_{\rm tag}$ is tagged as a $\bar{B}^0$ (signal-side: \Bz) and the bottom panel for candidates where $B_{\rm tag}$ is tagged as a $B^0$ (signal-side: \Bzb). The distribution and fit are integrated over \r-bin in the good tag region $0.25 \le r \le 1$ and in the signal region (left panel: $-0.16 < \Delta E < 0.08$ GeV, right panel: $M_{\rm bc} > 5.27$\,GeV/$c^2$).}
\label{fig:Flv_specific_62.8}
\end{figure}

Unlike $\Bz \to K^{+} \pim$ or $\Bp \to K^{+} \piz$, the kaon charge does not tag the $B$ flavor and therefore the $b$-quark flavor must be determined from the inclusive properties of the rest of the event (``flavor tagging'').
We measure the \CP-violating asymmetry of the \CP-eigenstate $\Bz \to \Kz \piz$ with the signal-side quark flavor \q, using the flavor content of the other $B$ meson ($B_{\rm tag}$) provided by the category-based flavor tagger~\cite{Abudinen:2018}. 
On an event-by-event basis, the tagging algorithm provides the flavor and associated dilution (\r), which is the complement to twice the mistag probability \w and measures the dilution of the asymmetry amplitude due to an incorrect tag (``wrong tag'') assignment. Events are categorized in 7 \r-bin intervals. For each \r-bin interval, the wrong tag fractions (\wr)
and tagging efficiencies (\er) are determined in control samples reconstructed in 2019 Belle II data~\cite{Abudinen:2020ozw} and constrained using Gaussian likelihoods in our fit. 
Systematic uncertainties are associated by varying
 the parameters from the control decay mode in the fit to the signal decay.
The asymmetry \akzpiz is determined from  a simultaneous maximum-likelihood fit to the unbinned \mbc-\de-\q distributions with signal-to-background fractions constrained by the yield fit of Sec.~\ref{sec:yields}.  The signal probability density function (PDF) of \q is the integral of the known $B^0 \to \Kz \piz$ decay-time evolution~\cite{Fujikawa_2010}

\begin{equation}
\begin{aligned}
\mathcal{P}_{sig}(q) &= \frac{1}{2}(1 + q \cdot (1-2\wr) \cdot (1-2\chi_d) \akzpiz) & \quad \\
\label{eq:q_Belle2_moriond}
\end{aligned}
\end{equation}
with the time-integrated mixing parameter $\chi_d$ set to its known value $\chi_d = 0.1858 \pm 0.0011$~\cite{PDG}.
We assume the background from charmless $B$ decays to be flavor symmetric as well as the continuum sample.
The resulting asymmetry is $\akzpiz = -0.40_{-0.44}^{+0.46}$, where the uncertainty includes only the statistical contribution.
In Fig.~\ref{fig:Flv_specific_62.8}, the results of the fit on well-tagged events  are displayed separately in \mbc and \de projections.

\section{Systematic uncertainties}
\label{sec:syst}

We consider several sources of systematic uncertainties. We assume the sources to be independent and add in quadrature the corresponding uncertainties.
Summaries of systematic uncertainties are reported in Tables~\ref{tab:SystematicsBF_overview} and \ref{tab:SystematicsAcp_overview} for the measurement of the branching fraction and \CP asymmetry, respectively. Each contribution is described in detail below.

\subsection{Systematic uncertainties for branching fraction measurement}
\subsection*{Tracking efficiency} 
We assess a systematic uncertainty associated with possible data-simulation discrepancies in the reconstruction of charged particles~\cite{Bertacchi:2020eez}.
The tracking efficiency in data agrees with the value observed in simulation within a $0.91\%$ uncertainty, which we (linearly) add as a systematic uncertainty for each final-state charged particle.

\subsection*{\PKzS~reconstruction efficiency} 
A small decrease, approximately linear with flight length, in the $\PKzS$ reconstruction efficiency was observed in early Belle~II data with respect to simulation.
We assess a systematic uncertainty based on dedicated studies performed for the $B \to \phi K^{(*)}$~analysis~\cite{Ale:2020}. We apply an uncertainty of $0.31\%$ for each centimeter of average flight length of the $\PKzS$~candidate, with an additional $15\%$ uncertainty accounting for the data and simulation mismatch between the second and third layer of the vertex detector,
resulting in a 3.82\% total systematic uncertainty. 

\subsection*{\Pgpz~reconstruction efficiency} 
We assess a systematic uncertainty associated with possible data-simulation discrepancies in the $\Pgpz$ reconstruction and selection using the decays \mbox{$B^0 \to D^{*-}(\to \overline{D}^0 (\to K^+ \pi^- \pi^0)\, \pi^-)\, \pi^+$} and \mbox{$B^0 \to D^{*-}(\to \overline{D}^0 (\to K^+ \pi^-)\, \pi^-)\, \pi^+$} where the selection of charged particles is identical and all distributions are weighted so as the $\pi^0$ momentum matches the $\pi^0$ momentum in charmless channels. We compare the yields obtained from fits to the $\Delta E$ distribution of reconstructed $\PB$~candidates
and obtain the \piz reconstruction efficiency in data and simulation. 
The efficiency ratio $\varepsilon_{\rm MC} (\piz) / \varepsilon_{\rm data} (\piz)$ is close to 1, and the 13.0\% uncertainty of the efficiency ratio is quoted as the systematic uncertainty.

\subsection*{Continuum-suppression efficiency} 
We evaluate possible data-simulation discrepancies in the continuum-suppression distributions using the control channel \mbox{$\PBplus\to\APDzero(\to\PKp\Pgpm\Pgpz)\,\Pgpp$}.
The selection efficiency obtained in data and simulation agrees well, hence the statistical uncertainty on data efficiency is assigned as the systematic uncertainty.

\subsection*{Number of $\PB\APB$~pairs} 
We  assign  a  $1.4\%$  systematic  uncertainty on the number of $\PB\APB$~pairs,  which  includes  the uncertainty  on  cross section,  integrated  luminosity~\cite{Abudinen:2019osb}, and  potential  shifts  from  the peak center-of-mass energy during the run periods.

\subsection*{Fit modelling}
Because we used empirical fit models for signal, we assess a systematic uncertainty associated with the choice of model. 
We consider the effect by performing independent variations on the \mbc and \de shapes of signal PDF modeling.
The uncertainty due to signal modeling is given by changing the nominal fit configuration to an alternate configuration, with the difference quoted as the systematic uncertainty. 
For both variations on \mbc and \de, around a 0.01\% uncertainty is obtained.
We perform a variation of the continuum background modelling, allowing for more degrees of freedom in the PDF. The resulting systematic uncertainty is about $1\%$.

\subsection{Systematic uncertainties for direct \CP asymmetry measurement}


\subsection*{Flavor tagging modelling}
A potential bias exists in flavor parameters between $\Bz \to \KS \piz$ and the control sample used in determining the flavor parameters. 
The uncertainty is assessed from the shift observed in the value of \akzpiz when fitting simulated data using flavor tagging parameters determined either from the signal decay or from the control decay.

\subsection*{\Bz mixing parameter $\chi_d$}
In the nominal fit, the value of the mixing parameter $\chi_d$ is fixed. 
Two alternative fits, where the fixed value of $\chi_d$ is increased and decreased by its uncertainty, are performed. The largest variation of \akzpiz from its known value is quoted as the systematic uncertainty.

\subsection*{$B$-decay background asymmetry}
In the nominal fit, the peaking backgrounds from charmless $B$ decays are assumed to be \CP symmetric. Since several decays with poorly known \CP-violating asymmetries may contribute, we perform two alternative fits, where the asymmetry of this background is fixed to $\pm 1$.  
We then assign as a systematic the resulting largest shift observed in \akzpiz.

\subsection*{Continuum background asymmetry}
In addition, the continuum background is assumed to be symmetric in the nominal fit. We allow for an asymmetry of this component in an alternative fit, and found that its value is consistent with zero with a $7\%$ uncertainty. We assign as a systematic uncertainty the observed shift on \akzpiz from this alternative fit.  


\begin{table}[h]
\centering
\footnotesize
\caption{Summary of the (fractional) systematic uncertainties of the branching fraction measurement.}
\begin{tabular}{l c c c c c c c c}
\hline\hline
 Source 		& $\delta \mathcal{B} (\%)$ \\
\hline
 Tracking efficiency & 1.8  \\
 $K_{S}^{0}$ reconstruction efficiency & 3.8  \\
 $\pi^{0}$ reconstruction efficiency & 13.0  \\
 Continuum-suppression efficiency & 2.4  \\
 N($B\bar{B}$) (as written in Eq.~\ref{eq:BF}) & 1.4  \\
 Signal model & $<$0.1  \\
 Continuum background model & 1.4  \\
\hline
 Total & 14.0  \\
\hline\hline
\end{tabular} 
\label{tab:SystematicsBF_overview}
\end{table}

\begin{table}[h]
\centering
\footnotesize
\caption{Summary of (absolute) systematic uncertainties in the \akzpiz measurement.}
\begin{tabular}{l c c c c c c}
\hline\hline
 Source 		& $\delta \akzpiz$ \\
\hline
 Flavor tagging modelling & 0.03  \\
 $B^0$ mixing parameter $\chi_d$ & $<$0.01  \\
 $B$-decay background asymmetry & 0.03  \\
 Continuum background asymmetry & 0.01  \\
\hline
 Total & 0.04  \\
\hline\hline
\end{tabular} 
\label{tab:SystematicsAcp_overview}
\end{table}

\section{Result and future impact}
\label{sec:summary}
We report a measurement of 
the direct \CP asymmetry of the $\Bz \to \Kz \piz$ decay and an update of its branching fraction that supersedes the value measured in Ref.~\cite{belleiicollaboration2020measurements}.
We use data collected by Belle~II experiment in 2019 and 2020, corresponding to $62.8\,\si{fb^{-1}}$ of integrated luminosity, collected at the $\Upsilon(4S)$ resonance. We used simulation to devise and optimize the candidate selection. 
We reconstruct and select $45^{+9}_{-8}$ signal candidates, from which we determine the \CP asymmetry and the branching fraction to be

\begin{center}
$\akzpiz = -0.40_{-0.44}^{+0.46} (\rm stat) \pm 0.04 (\rm syst)$, and 
\end{center}
\begin{center}
$\mathcal{B}(\Bz \to \Kz \piz) = [8.5_{-1.6}^{+1.7} (\rm stat)  \pm 1.2  (\rm syst)] \times 10^{-6}$.
\end{center}
The value of the branching fraction is in agreement with known determinations. 
The direct \CP asymmetry \akzpiz is measured for the first time at Belle~II, paving the way
for a precise test of the isospin sum rule with larger Belle II datasets.


Belle II plays a crucial role in updating the isospin sum rule results, as it is the only running experiment that reports measurements of branching fraction and \CP-violating asymmetries of $\Bz \to \Kz\piz$ decays. We investigate the impact that these and future measurements will have on the testing power of the isospin sum rule. As a figure of merit, we use the expected resolution on \ikpi (sensitivity) determined by extrapolating the current measurements under certain assumptions and approximations. 
The statistical uncertainties in the inputs for \ikpi are assumed to scale as $1/\sqrt{\int\mathcal{L} dt}$ and the corresponding systematics are assumed to be less than or equal to the statistical uncertainties.
For simplicity, we symmetrize asymmetric uncertainties and treat each measurement as an independent input, when incorporating it into the calculation of the \ikpi. 
In the calculation, we adopt a simple weighted average for the combination of new Belle II inputs with the current world average~\cite{PDG}.

\begin{figure}[h!]
 \centering
    \centering
    \subfigure{\includegraphics[width=0.9\textwidth]{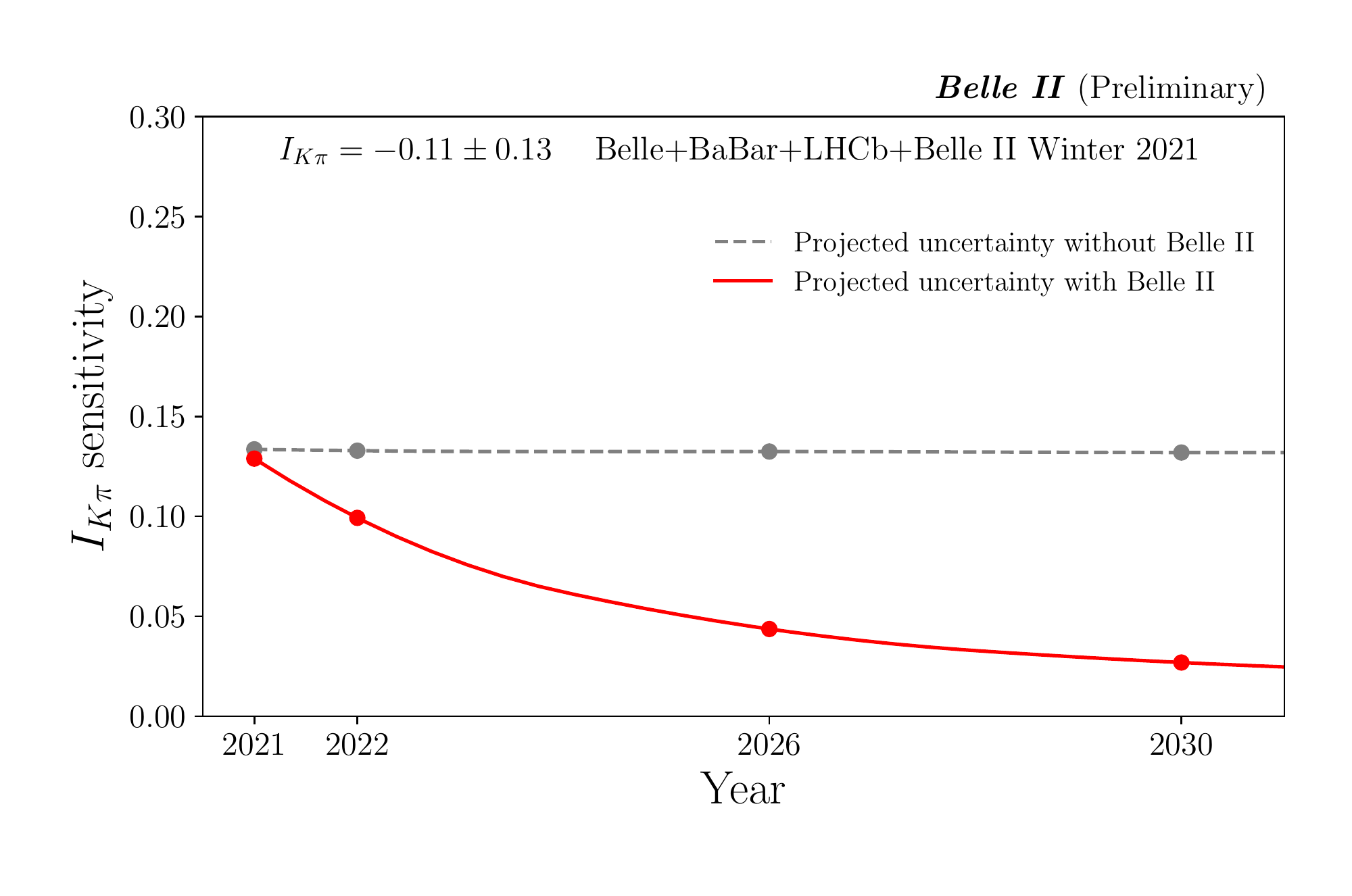}}
 \caption{The projected uncertainty on \ikpi with and without Belle II inputs. The inputs for \ikpi are averages of the estimated updates from ongoing LHCb and Belle II experiments with current world averages~\cite{PDG}. The red curve shows a projection when updates on the complete set of $K\pi$ measurements are considered, and the grey curve is the case if only $A_{K^+\pi^-}, A_{K^+\pi^0}, A_{K^0\pi^+}$ are updated by LHCb. The projection corresponds to the luminosity plans from LHCb and Belle II.}
 \label{fig:IKpiUncertYear}
\end{figure}

\begin{figure}[h!]
 \centering
    \centering
    \subfigure{\includegraphics[width=\textwidth]{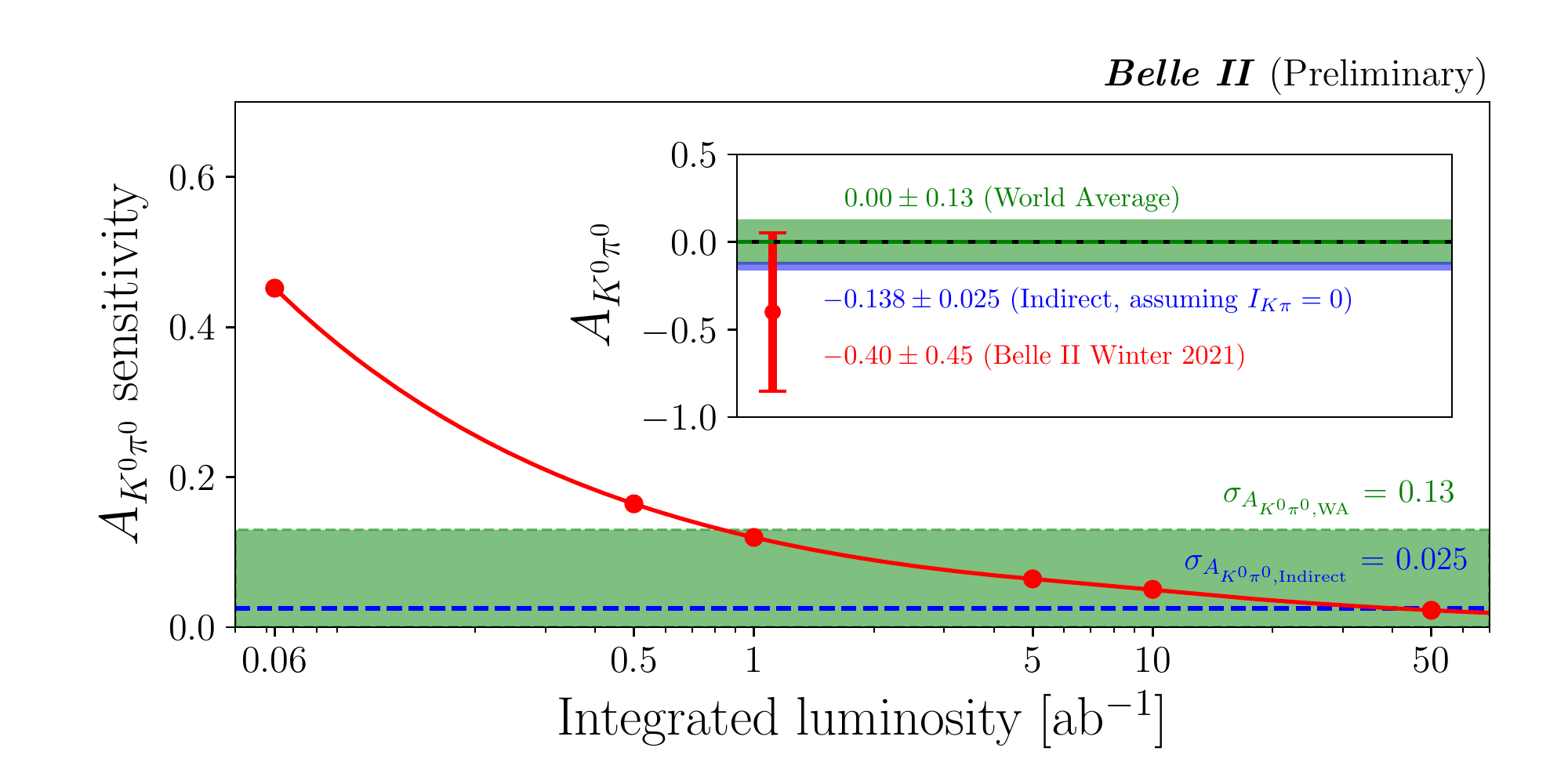}}
 \caption{The projected uncertainty on \akzpiz measurement.
 The inset panel shows the comparison of (red marker) the measurement reported here with (green band) the world average value, and (blue band) the indirect determination from Eq.~\ref{eq:SumRule} assuming \ikpi = 0 and world average values for the other inputs. 
 The red curve in the main panel is Belle II's expected uncertainty on the \akzpiz measurement as a function of the integrated luminosity, while the green and blue dashed lines are the uncertainties of the world average value and of the indirect determination, respectively.}
 \label{fig:AK0pi0Uncert}
\end{figure}

Fig.~\ref{fig:IKpiUncertYear} shows a projection of the \ikpi sensitivity in the next decade according to the current luminosity plans from LHCb~\cite{Bediaga:2018lhg} and the Belle II  experiment.
For measurements of $A_{K^+\pi^-}, A_{K^+\pi^0}, A_{K^0\pi^+}$, we average previous Belle and BaBar results with projected inputs from LHCb, which are anticipated to dominate the precision on these quantities and are scaled with expected sample size.
We compare the sensitivity on \ikpi in the scenarios with and without Belle II contributions. As expected, the sensitivity to \ikpi is mostly determined by the $\Kz\piz$ inputs, which makes Belle II's $\Kz\piz$ measurements essential.
We provide an estimation for the future sensitivity of Belle II's \akzpiz measurement in Fig.~\ref{fig:AK0pi0Uncert}.

The projection based on current inputs is conservative. We expect a gain in the signal efficiency and the overall sensitivity by relaxing the continuum suppression requirement and including the continuum suppression variable in the fit.
A measurement using a time-dependent \CP violation method with a specific optimization on the vertex quality (resolution) requirement can also bring additional benefits.

\section*{Acknowledgments}

We thank the SuperKEKB group for the excellent operation of the
accelerator; the KEK cryogenics group for the efficient
operation of the solenoid; the KEK computer group for
on-site computing support; and the raw-data centers at
BNL, DESY, GridKa, IN2P3, and INFN for off-site computing support.
This work was supported by the following funding sources:
Science Committee of the Republic of Armenia Grant No. 20TTCG-1C010;
Australian Research Council and research grant Nos.
DP180102629, 
DP170102389, 
DP170102204, 
DP150103061, 
FT130100303, 
FT130100018,
and
FT120100745;
Austrian Federal Ministry of Education, Science and Research,
Austrian Science Fund No. P 31361-N36, and
Horizon 2020 ERC Starting Grant no. 947006 ``InterLeptons''; 
Natural Sciences and Engineering Research Council of Canada, Compute Canada and CANARIE;
Chinese Academy of Sciences and research grant No. QYZDJ-SSW-SLH011,
National Natural Science Foundation of China and research grant Nos.
11521505,
11575017,
11675166,
11761141009,
11705209,
and
11975076,
LiaoNing Revitalization Talents Program under contract No. XLYC1807135,
Shanghai Municipal Science and Technology Committee under contract No. 19ZR1403000,
Shanghai Pujiang Program under Grant No. 18PJ1401000,
and the CAS Center for Excellence in Particle Physics (CCEPP);
the Ministry of Education, Youth and Sports of the Czech Republic under Contract No.~LTT17020 and 
Charles University grants SVV 260448 and GAUK 404316;
European Research Council, 7th Framework PIEF-GA-2013-622527, 
Horizon 2020 ERC-Advanced Grants No. 267104 and 884719,
Horizon 2020 ERC-Consolidator Grant No. 819127,
Horizon 2020 Marie Sklodowska-Curie grant agreement No. 700525 `NIOBE,' 
and
Horizon 2020 Marie Sklodowska-Curie RISE project JENNIFER2 grant agreement No. 822070 (European grants);
L'Institut National de Physique Nucl\'{e}aire et de Physique des Particules (IN2P3) du CNRS (France);
BMBF, DFG, HGF, MPG, and AvH Foundation (Germany);
Department of Atomic Energy under Project Identification No. RTI 4002 and Department of Science and Technology (India);
Israel Science Foundation grant No. 2476/17,
United States-Israel Binational Science Foundation grant No. 2016113, and
Israel Ministry of Science grant No. 3-16543;
Istituto Nazionale di Fisica Nucleare and the research grants BELLE2;
Japan Society for the Promotion of Science,  Grant-in-Aid for Scientific Research grant Nos.
16H03968, 
16H03993, 
16H06492,
16K05323, 
17H01133, 
17H05405, 
18K03621, 
18H03710, 
18H05226,
19H00682, 
26220706,
and
26400255,
the National Institute of Informatics, and Science Information NETwork 5 (SINET5), 
and
the Ministry of Education, Culture, Sports, Science, and Technology (MEXT) of Japan;  
National Research Foundation (NRF) of Korea Grant Nos.
2016R1\-D1A1B\-01010135,
2016R1\-D1A1B\-02012900,
2018R1\-A2B\-3003643,
2018R1\-A6A1A\-06024970,
2018R1\-D1A1B\-07047294,
2019K1\-A3A7A\-09033840,
and
2019R1\-I1A3A\-01058933,
Radiation Science Research Institute,
Foreign Large-size Research Facility Application Supporting project,
the Global Science Experimental Data Hub Center of the Korea Institute of Science and Technology Information
and
KREONET/GLORIAD;
Universiti Malaya RU grant, Akademi Sains Malaysia and Ministry of Education Malaysia;
Frontiers of Science Program contracts
FOINS-296,
CB-221329,
CB-236394,
CB-254409,
and
CB-180023, and SEP-CINVESTAV research grant 237 (Mexico);
the Polish Ministry of Science and Higher Education and the National Science Center;
the Ministry of Science and Higher Education of the Russian Federation,
Agreement 14.W03.31.0026;
University of Tabuk research grants
S-0256-1438 and S-0280-1439 (Saudi Arabia);
Slovenian Research Agency and research grant Nos.
J1-9124
and
P1-0135; 
Agencia Estatal de Investigacion, Spain grant Nos.
FPA2014-55613-P
and
FPA2017-84445-P,
and
CIDEGENT/2018/020 of Generalitat Valenciana;
Ministry of Science and Technology and research grant Nos.
MOST106-2112-M-002-005-MY3
and
MOST107-2119-M-002-035-MY3, 
and the Ministry of Education (Taiwan);
Thailand Center of Excellence in Physics;
TUBITAK ULAKBIM (Turkey);
Ministry of Education and Science of Ukraine;
the US National Science Foundation and research grant Nos.
PHY-1807007 
and
PHY-1913789, 
and the US Department of Energy and research grant Nos.
DE-AC06-76RLO1830, 
DE-SC0007983, 
DE-SC0009824, 
DE-SC0009973, 
DE-SC0010073, 
DE-SC0010118, 
DE-SC0010504, 
DE-SC0011784, 
DE-SC0012704, 
DE-SC0021274; 
and
the Vietnam Academy of Science and Technology (VAST) under grant DL0000.05/21-23.

\bibliography{belle2}
\bibliographystyle{belle2-note}

\end{document}